\documentclass[twocolumn]{IEEEtran}
\usepackage[noadjust]{cite}
\usepackage[cmex10]{amsmath}
\usepackage{amsfonts}
\usepackage{amssymb}
\usepackage{algorithm}
\usepackage{algorithmic}
\usepackage[dvipsnames]{xcolor}
\usepackage[colorlinks,linkcolor=black,anchorcolor=black,citecolor=black]{hyperref}
\usepackage{graphicx}
\usepackage{verbatim}
\usepackage{indentfirst}
\usepackage{epsfig}
\usepackage{bm}
\usepackage{multirow}
\usepackage{subfigure}
\usepackage{setspace}

\IEEEoverridecommandlockouts
\newtheorem{remark}{\underline{Remark}}[section]

\newcommand{\mv}[1]{\mbox{\boldmath{$ #1 $}}}

\begin{document}
\title{Passive Reflection Codebook Design for IRS-Integrated Access Point }
\author{Yuwei Huang, {\it Student Member, IEEE}, Lipeng Zhu, {\it Member, IEEE}, and Rui Zhang, {\it Fellow, IEEE}

\thanks{

Y. Huang is with the NUS Graduate School, National University of Singapore, Singapore 119077, and also with the Department of Electrical and Computer Engineering, National University of Singapore, Singapore 117583 (e-mail: yuweihuang@u.nus.edu).

L. Zhu is with the Department of Electrical and Computer Engineering, National University of Singapore, Singapore 117583 (e-mail:zhulp@nus.edu.sg).

R. Zhang is with School of Science and Engineering, Shenzhen Research Institute of Big Data, The Chinese University of Hong Kong, Shenzhen, Guangdong 518172, China (e-mail: rzhang@cuhk.edu.cn). He is also with the Department of Electrical and Computer Engineering, National University of Singapore, Singapore 117583 (e-mail: elezhang@nus.edu.sg).
}
}
\maketitle

\begin{abstract}
Intelligent reflecting surface (IRS) has emerged as a promising technique to control wireless propagation wireless environment for improving the communication performance cost-effectively and extending the wireless signal coverage of access point (AP). In order to reduce the path-loss of the cascaded user-IRS-AP channels, the IRS-integrated AP architecture has been proposed to deploy the antenna array of the AP and the IRSs within the same antenna radome. To reduce the pilot overhead for estimating all IRS-involved channels, in this paper, we propose a novel codebook-based IRS reflection design for the IRS-integrated AP to enhance the coverage performance in a given area. In particular, the codebook consisting of a small number of codewords is designed offline by employing an efficient sector division strategy based on the azimuth angle. To ensure the performance of each sector, we optimize its corresponding codeword for IRS reflection pattern to maximize the sector-min-average-effective-channel-power (SMAECP) by applying the alternating optimization (AO) and semidefinite relaxation (SDR) methods. With the designed codebook, the AP performs the IRS reflection training by sequentially applying all codewords and selects the one achieving the best communication performance for data transmission. Numerical results show that our proposed codebook design can enhance the average channel power of the whole coverage area, as compared to the system without IRS. Moreover, the proposed codebook-based IRS reflection design is compared with several benchmark schemes, which achieves significant performance gain in both single-user and multi-user transmissions.
\end{abstract}

\section{Introduction}

With the recent progress in digitally-controlled metasurfaces, intelligent reflecting surface (IRS) has emerged as  as economically efficient method to create intelligent and adaptable radio environments, catering to the needs of next-generation wireless communication systems \cite{tutorial}. Specifically, IRS consists of a massive number of passive reflecting elements, which are able to tune the amplitudes and/or phase shifts of incident signals in real time, thereby enabling dynamic control over the wireless signal propagation environment. Thus, IRS can be applied in wireless communication systems to achieve assorted purposes, such as passive beamforming, interference nulling/cancellation, channel distribution enhancement, etc \cite{magazine,tutorial_new}. Due to its passive nature, IRS eliminates the need for radio frequency (RF) chains, resulting in significantly reduced hardware costs and energy consumption. As a result, IRS is considered as a promising candidate for the six-generation (6G) wireless communication systems  \cite{6G,6G2}, which can achieve a quantum-leap improvement in capacity and energy efficiency over today's wireless systems. Owing to the great potential of IRS, it has been extensively investigated for various wireless systems and applications, such as non-orthogonal multiple access (NOMA)  \cite{noma,noma1}, orthogonal frequency division multiplexing (OFDM) \cite{ofdm,ofdm1}, secrecy communication \cite{security,security1}, mobile edge computing (MEC) \cite{MEC,MEC1}, multiple-input multiple-output (MIMO) \cite{multi_cell,cell1}, unmanned aerial vehicle (UAV)-ground communications \cite{uav,uav1}, multi-antenna communication \cite{antenna,antenna1}, relaying communication \cite{relay,relay1}, and so on.

In practice, it is essential to ensure the proper deployment of IRSs between the users and base station (BS) (or access point (AP)). The deployment of IRSs should reduce the path loss resulting from the product of distances along the cascaded user-IRS-BS (or AP) channels \cite{tutorial}. Most of the existing works considered to deploy IRSs close to user terminals (e.g., at hotspot, cell edge, and on moving vehicles) to improve their communication rates. For example, the authors in \cite{qingqing} proposed to deploy the IRS near the user cluster, where the transmit beamforming at the BS and the reflect beamforming of the IRS were jointly optimized to minimize the total transmit power at the BS, and the power scaling law with the number of reflecting elements at the IRS was derived. In \cite{multi_cell}, the authors invoked an IRS at the boundary of multiple cells to assist the downlink transmission to cell-edge users and mitigate the inter-cell interference, where the active precoding matrices at the BSs and the phase shifts at the IRS were jointly optimized to maximize the weighted sum-rate of all users. In contrast, a novel IRS-empowered BS architecture was introduced in \cite{my1} to deploy multiple IRSs in proximity to the BS. In this work, a novel approach was proposed to lower the overhead of channel estimation by selecting specific portions of the cascaded channels for estimation purposes, and a transmission protocol including two-stage was designed to realize the efficient process of IRS channel estimation and user data transmission. Besides, the authors in \cite{DRL,machine_learning} deployed the IRS near the BS and adopted the deep reinforcement learning (DRL) technique to design the passive reflection of the IRS for assisting the communication between the BS and multiple users. To further enhance the system performance, an innovative strategy was introduced in \cite{deploy} for hybrid deployment to harness the complementary strengths of both user- and BS-side IRSs, where the additional inter-IRS reflection link can be exploited to achieve higher achievable rate of the users \cite{double1,double2}. Nevertheless, the above strategies all deploy IRSs within an environment, where the separation distance between the IRS and its connected user or BS (or AP) remains considerable, typically exceeding several hundred carrier wavelengths. In such scenarios, the substantial path loss resulting from the product-distance of the cascaded user-IRS-BS (or AP) channels potentially undermines the performance benefits offered by the IRSs, and the signaling overhead between the IRSs and the BS (or AP) for channel estimation and remote control remains a challenging issue.

In order to address the aforementioned issues, a novel architecture, termed the {\it``IRS-integrated BS (or AP)''}, has been proposed in \cite{my}. This architecture involves co-locating BS's  (or AP's) antenna array and IRSs within the same antenna radome to serve users. Consequently, the separation distance between the BS's (or AP's) antenna array and IRSs can be significantly reduced, which typically ranges from several to tens of wavelengths. This substantially mitigates the path loss associated with IRS reflection channels and also diminishes the signaling overhead between the IRSs and BS (or AP). It is worth noting that the key distinction of the IRS-integrated BS (or AP) architecture from the other existing approaches for deploying IRS at the BS (or AP), such as active holographic MIMO surface \cite{hMIMO1,hMIMO2}, dynamic metasurface antenna \cite{DAM1,DAM2}, receiving IRS \cite{rIRS}, and IRS for sensing \cite{sensing}, lies in that these existing methods require embedding RF chains and signal processing units on IRS to manipulate its analog beampattern for transmission/reception. In contrast, our proposed IRS-integrated BS (or AP) incorporates only passive reflecting elements into the BS's (or AP's) antenna radome. This new architecture enables the reconfiguration of electromagnetic propagation to/from the BS (or AP) to enhance communication/sensing performance, but without any modifications to the RF front-end structure of existing BS (or AP) antennas. Consequently, our proposed IRS-integrated BS (or AP) requires lower energy consumption and incurs lower hardware cost, which is also more compatible with the existing BS (or AP) in wireless systems. Furthermore, a novel stacked intelligent metasurface (SIM) technology has been recently proposed in \cite{SIM1,SIM2,SIM3} to deploy {\it multi-layer metasurfaces} in front of the antenna array of the BS (or AP) for achieving transmit precoding and receive combining in the electromagnetic wave domain. In contrast, our proposed IRS-integrated BS (or AP) deploys multiple IRSs with {\it a single-layer metasurface} surrounding the BS's (or AP's) antenna array, where the passive reflection of the IRSs is designed together with the BS's (or AP's) active beamforming to improve the system performance.

For the IRS-integrated BS (or AP), a practically important problem is how to design the IRS reflection based on the available channel state information (CSI) of the users for improving their communication performance. Towards this end, two solutions have been proposed in \cite{my}. First, the element-wise channel model was adopted to construct the singe-reflection and double-reflection channels in terms of the angle-of-arrivals (AoAs) at the BS (or AP) and the complex coefficients of the incident paths. Then, the AoAs and complex coefficients of all paths from each user can be estimated by the antenna array at the BS (or AP) by turning off all IRSs. Based on them, the effective channel of each user with IRS reflection turned on can be derived for IRS reflection design. However, this method needs to acquire the knowledge of the AoAs and complex coefficients of all channel paths from each user (instead of its channel only), which is thus more difficult to be implemented compared to traditional channel estimation methods for IRS-assisted communications \cite{channel_estimation0,channel_estimation2,liuliang,anchor,double_estimation,group,sparsity,channel_estimation_survey}. Second, an iterative random phase maximization algorithm (IRPA) was proposed to find a suboptimal IRS reflection solution without the need to explicitly estimate the CSI of any user. Specifically, a given number of IRS training reflection patterns were randomly generated, and then the reflection pattern achieving the best communication performance was selected as the one for data transmission. Despite of its simplicity for implementation, it was shown in \cite{my} that IRPA usually needs a sufficiently long training time for achieving a good communication performance, which may be inefficient for data transmission.

\begin{figure*} 
	\centering
	\subfigure{
\begin{minipage}[t]{0.48\textwidth}
		\includegraphics[width=7cm]{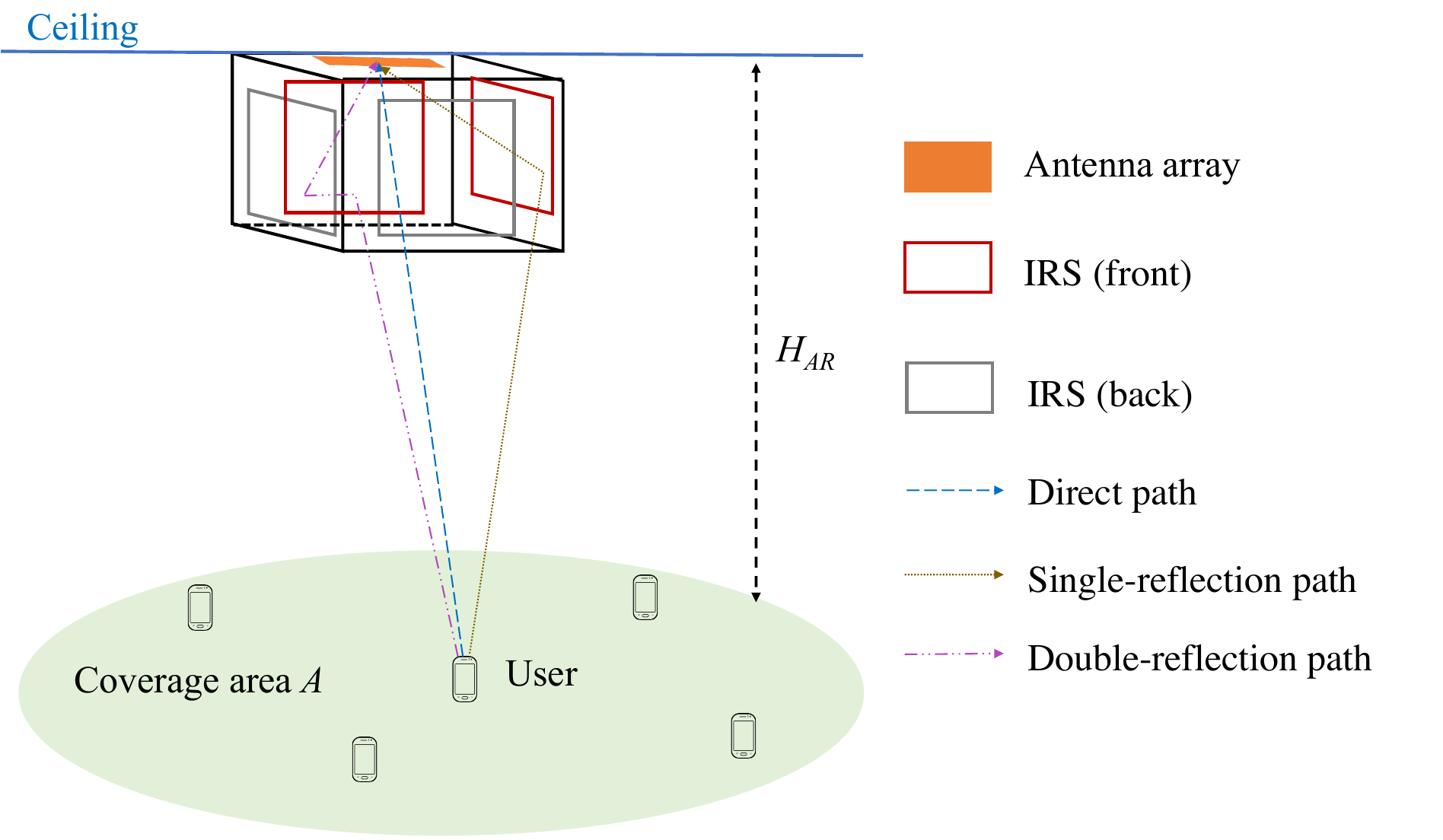}
		\end{minipage}}
	\subfigure{
\begin{minipage}[t]{0.48\textwidth}
		\includegraphics[width=7cm]{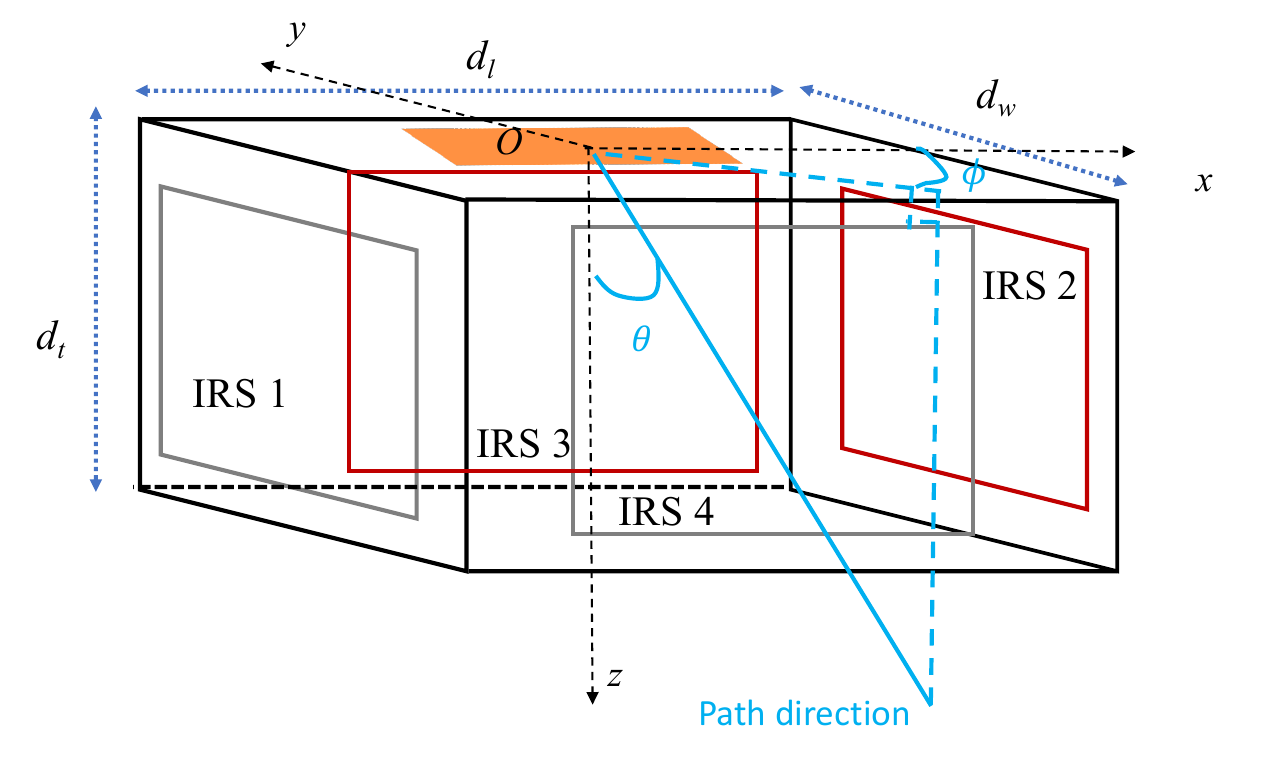}
		\end{minipage}}
		\caption{System model and architecture of the IRS-integrated AP.}
	\label{system_model} 
\end{figure*}
To avoid the estimation of the AoAs and complex coefficients of all incident paths, as well as reduce the channel training overhead, in this paper, we propose a novel codebook-based IRS reflection design for the IRS-integrated AP (IRS-AP), which is installed on the ceiling to communicate with the users randomly distributed in a given coverage area on the ground, as shown in Fig. \ref{system_model}. Specifically, a codebook with a small number of IRS reflection patterns/codewords is first designed offline without requiring CSI with users and stored at the IRS-AP. Then, the users send pilot signals to the IRS-AP for effective channel estimation, while in the meanwhile, the IRS-AP evaluates the communication performance according to the estimated channels under different IRS reflection patterns/codewords. Finally, the IRS reflection pattern/codeword achieving the best communication performance is selected for data communication. The main contributions of this paper are summarized as follows.
\begin{itemize}
\item First, the codebook for IRS reflection is designed to enhance the average channel power of the whole coverage area. To achieve a small codebook size for reducing the channel training overhead, an efficient sector division strategy is proposed to divide the whole coverage area into several non-overlapped sectors based on the azimuth angle. Then, for single-user transmission, the number of sectors is set as the given codebook size to maximize the user's effective channel power for improving its communication performance. While for the multi-user transmission, the overall codebook is constructed as the union of all IRS codebooks with different numbers of sectors, in order to cater to users at different locations.
\item Next, the IRS reflection pattern/codeword corresponding to each sector is designed to improve its worst-case performance, which is the average effective channel power over all azimuth angles in this sector by fixing the elevation angle as its largest value and thus defined as {\it sector-min-average-effective-channel-power (SMAECP)}. Specifically, the IRS reflection pattern/codeword is optimized to  maximize the SMAECP of the corresponding sector, subject to the unit-modulus constraints of all IRS reflecting elements. The formulated problem is transformed to an approximated problem by discretizing the continuous azimuth angles, which is then efficiently solved by applying the alternating optimization (AO) and semidefinite relaxation (SDR) methods.
\item Finally, numerical results are provided to validate that the proposed codebook design can enhance the average channel power of the whole coverage area, as compared to the system without IRS. Besides, the proposed codebook-based IRS reflection design is compared with several benchmark schemes, which achieves significant performance gain in both single-user and multi-user transmissions. In addition, it is shown that the proposed codebook-based IRS reflection design only requires slow adaptation to wireless channels for reducing the channel training overhead, where the IRS reflection pattern can remain unchanged for a long time owing to the sector division strategy adopted for codebook design. Furthermore, the optimal number of sectors for multi-user transmission is shown to vary with different user locations.
\end{itemize}

It is worth noting that there exist several works on reflection codebook design for IRS-assisted communications (see, e.g., \cite{codebook1,codebook3,codebook4,codebook5}). However, these works usually need large codebook size (e.g., hundreds to thousands) to achieve high IRS directional gain for ensuring the system performance. Besides, the corresponding IRS reflection pattern should be updated frequently to track user channel changes, which incurs high channel training overhead in practice. In contrast, the size of the codebook designed in this paper is dramatically reduced based on the efficient sector division strategy, and the IRS reflection pattern can remain unchanged for a long time to achieve slow adaptation to wireless channels, both of which help reduce the channel training overhead for adjusting IRS reflection pattern. In addition, the conventional codebook designs consider IRS single reflection only, while our considered codebook design needs to take both IRS single reflection and double reflection into consideration due to multiple IRSs deployed inside the AP's antenna radome (see Fig. \ref{system_model}).

The rest of this paper is organized as follows. Section II presents the system and channel models of the IRS-integrated AP. Section III presents the procedures for reflection codebook design based on sector division. Section IV presents the proposed solution to the formulated problem for codeword design. Section V presents numerical results to verify the efficacy of our proposed codebook-based IRS reflection design. Finally, Section VI gives the conclusion of this paper.

{\it Notation:} In this paper, italic, bold-face lower-case and bold-face upper-case letters denote scalars, vectors, and matrices, respectively. For a matrix $\mv A$, its transpose, conjugate transpose, the determinant, the trace and the rank are denoted as $\mv A^{T}$, $\mv A^{H}$, $\det(\mv A)$, $\text{trace}(\mv A)$, and $\text{rank}(\mv A)$, respectively. $[\mv A]_{i,j}$ denotes the $(i,j)$-th entry of the matrix $\mv A$. For a vector $\mv a$,  the norm is denoted as $\lVert \mv a \rVert$, and $\text{diag}(\mv a)$ denotes a square diagonal matrix with the elements of $\mv a$ on the main diagonal. For a complex number $s$, the conjugate and amplitude are respectively denoted as $s^{*}$ and $|s|$. $\mathbb{R}^{x\times y}$ denotes the space of $x\times y$ real matrices. $\mathbb{C}^{x\times y}$ denotes the space of $x\times y$ complex matrices. The distribution of a circularly symmetric complex Gaussian (CSCG) random variable with mean zero and variance $\sigma^{2}$ is denoted by $\mathcal{CN}(0,\sigma^{2})$, and $\sim$ stands for ``distributed as''. $\mv I_{M}$ denotes an identity matrix with its dimension of $M$. $i$ denotes the imaginary unit, i.e., $i^2=-1$. $\otimes$ represents the Kronecker product. $\mathcal{O}(\cdot)$ denotes the the big O notation. $\lfloor b \rfloor$ denotes the floor of a real number $b$.

\section{IRS-Integrated AP: System and Channel Models}
\subsection{System Model}
As shown in Fig. \ref{system_model}, we consider an IRS-AP installed on the ceiling\footnote{Although we consider the ceiling-mounted AP shown in Fig. \ref{system_model} for ease of exposition, our proposed design in this paper is applicable to other AP or BS deployment scenarios \cite{my}, e.g., with different suspending/tilt angles for the antenna radome.} of hight $H_{\text{AR}}$ from the ground to communicate with users at arbitrary locations in a given coverage area, denoted by $\mathcal A$. In particular, the IRS-AP deploys an (active) antenna array and $J=4$ (passive) IRSs with distinct orientations within the same cuboid-shaped antenna radome, where $\mathcal J\triangleq\{1,2,3,4\}$ denotes the set of all IRSs. The antenna array is deployed at the center of the top surface of the antenna radome facing to the ground, while IRSs 1--4 are strategically positioned on the left, right, back, and front side-faces of the antenna radome, with each IRS oriented perpendicular to the antenna array. The length, width, and thickness of the antenna radome are respectively denoted as $d_l$, $d_w$, and $d_t$. Without loss of generality, we consider a three-dimensional (3D) coordinate system for the IRS-AP. In this system, we set the origin $O$ as the antenna array's center, and put the antenna array at the $x$--$O$--$y$ plane. The uniform planar array (UPA) is adopted to model the antenna array, which consists of a total of $M=M_{x}\times M_{y}$ antennas, with $M_{x}$ and $M_{y}$ representing the number of antennas along the $x$ and $y$ axes, respectively. The spacing between adjacent antennas is denoted as $d_{A}$. Each IRS $j,~j\in\mathcal J$, is also modeled as a UPA of size $N_{j}=N_{j,1}\times N_{j,2}$, with $N_{j,1}$ and $N_{j,2}$ denoting the number of reflecting elements along axes $x$ (or $y$) and $z$, respectively.  The distance between two adjacent reflecting elements within each IRS is defined as $d_{I}$, and each reflecting element is configured with an aperture area of $A=\sqrt{A}\times\sqrt{A}$. For convenience, we denote the sets of all AP's antennas and all reflecting elements at IRS $j$ as $\mathcal M\triangleq\{1,2,\cdots,M\}$ and $\mathcal {N}_{j}\triangleq\{1,2,\cdots,N_j\},~j\in\mathcal J$, respectively. Notice that each IRS only has half-space reflection space, each IRS must be placed to face the AP's antenna array for effective signal reflection between them. Moreover, as shown in Fig. \ref{system_model}, let $\theta$ and $\phi$ denote the elevation and azimuth angles of a particular signal path arriving at the IRS-AP with respect to (w.r.t.) the AP's antenna array. Furthermore,  we model the coverage area $\mathcal A$ by the IRS-AP in terms of $(\theta,\phi)$ as $\mathcal{A}=\{(\theta,\phi)|\theta\in[0,\theta_{\max}],\phi\in[0,2\pi)\}$ with $\theta_{\max}$ ($0\leq \theta_{\max}\leq\frac{\pi}{2}$) denoting the maximum elevation angle, which is pre-determined based on the IRS-AP's coverage requirement.

\begin{remark}\label{deployment}
Based on the antenna radome's size (i.e., $d_{l}$, $d_{w}$, and $d_{t}$), and the coverage area's range $\mathcal A$ (i.e., $\theta_{\max}$), we can determine the maximum number of deployable reflecting elements at each IRS $j$, which is defined as $N_{j,\max}=N_{j,1,\max}\times N_{j,2,\max},~j\in\mathcal J$, with $N_{j,1,\max}$ and $N_{j,2,\max}$ representing the maximum number of reflecting elements that can be deployed along axes $x$ (or $y$) and $z$ of IRS $j$, respectively. To deploy more reflecting elements at each IRS, we set $\sqrt{A}=d_{I}$. It can be shown by elementary geometry that $N_{1,1,\max}=N_{2,1,\max}=\lfloor d_{w}/d_{I}\rfloor$, $N_{3,1,\max}=N_{4,1,\max}=\lfloor d_{l}/d_{I}\rfloor$, and $N_{j,2,\max}=\lfloor\min(\frac{d_{t}}{d_{I}},\frac{d_l}{d_I\tan\theta_{\max}},\frac{d_{w}}{d_{I}\tan\theta_{\max}})\rfloor$, where $\frac{d_l}{d_I\tan\theta_{\max}}$ and $\frac{d_{w}}{d_{I}\tan\theta_{\max}}$ are adopted to guarantee that the signals arriving from the reflection half-space of one IRS are not obstructed by its opposite side IRS. Next, when the number of reflecting elements at IRS $j$ is given as $N_{j}\leq N_{j,\max}$, the reflecting elements should be deployed to reduce the antenna-IRS distance and inter-IRS distance to minimize the path loss among them. To this end, we set $N_{j,2}=N_{j,2,\max}$ and $N_{j,1}=N_{j}/N_{j,2,\max},~j\in\mathcal J$, by first deploying reflecting elements along axis $z$, and then along axis $x$ (or $y$).
\end{remark}

\subsection{Channel Model}
In general, the effective channel between any location in the coverage area $\mathcal A$ and the IRS-AP is the superposition of the channel responses of multiple paths, where each path's the channel response is given by the product of its complex coefficient for the arriving uniform plane wave (UPW) (which is determined by the propagation environment) and the effective array response vector (EARV) at the IRS-AP. Moreover, the EARV at the IRS-AP for each signal path to the antenna array from a given AoA, i.e., $(\theta,\phi)$, is the superposition of the array response vectors corresponding to its three split signal paths arriving at the AP's antenna array, namely, the direct path without any IRS reflecting element's reflection, the single-reflection path with only one IRS reflecting element's reflection involved, and the double-reflection path with two IRS reflecting elements' successive reflections involved\footnote{Notice that more than two-hop IRS reflections may also exist in our proposed IRS-integrated AP architecture. However, their effect would be marginal to our proposed design based only on IRSs' single- and double-hop reflections. This is because no passive beamforming gain can be achieved via the more than two-hop reflections in our design to compensate the more severe multiplicative path loss.}, as shown in Fig. \ref{system_model}. In the following, we first model the direct, single- and double-reflection array response vectors in terms of the AoA $(\theta,\phi)$, and then derive their overall EARV at the IRS-AP for the path direction $(\theta,\phi)$. Based on them, we finally model the effective channel between any location in the coverage area $\mathcal{A}$ and the antenna array.

Let $\mv e(\bar{\phi},\bar{M})=[1,e^{i\pi\bar{\phi}},\cdots,e^{i(\bar{M}-1)\pi\bar{\phi}}]^{T}$ denote the one-dimensional (1D) steering vector function, with $\bar{M}$ representing the array size and $\bar{\phi}$ representing the steering angle. Thus, the direct array response vector at the AP's UPA for the path direction $(\theta,\phi)\in\mathcal{A}$, denoted as $\mv h_{d}(\theta,\phi)\in\mathbb{C}^{M\times 1}$, is given by
\begin{align}
\mv h_d(\theta,\phi)&=\sqrt{G_{A}(\theta,\phi)}\mv e\left(\frac{2d_{A}}{\lambda}\sin\theta\cos\phi,M_{x}\right)\nonumber\\
&\otimes \mv e\left(\frac{2d_{A}}{\lambda}\sin\theta\sin\phi,M_{y}\right),\label{direct}
\end{align}
with $\lambda$ denoting the carrier wavelength and $G_{A}(\theta,\phi)$ denoting the antenna gain of the AP corresponding to direction $(\theta,\phi)$.

According to \cite{my}, owing to the ultra-short distance between the antenna array and IRSs, as well as the fixed orientation of the IRSs, the single-reflection and double-reflection array response vectors can be directly expressed as the function of path direction $(\theta,\phi)$, both of which follow the near-field channel model. As a result, denote $\mv f_{j}^{n_{j}}(\theta,\phi)\in\mathbb{C}^{M\times 1}$ as the single-reflection array response vector for the path direction $(\theta,\phi)\in\mathcal{A}$ via the IRS $j$'s $n_j$-th reflecting element to the AP's antenna array with $j\in\mathcal J,~n_j\in\mathcal{N}_{j}$, and $\mv g_{j,q}^{n_j,n_q}(\theta,\phi)\in\mathbb{C}^{M\times1}$ as the double-reflection array response vector for the path direction $(\theta,\phi)\in\mathcal{A}$ via the IRS $j$'s $n_j$-th reflecting element and then the IRS $q$'s $n_q$-th reflecting element to the AP's antenna array with $q\neq j\in\mathcal J,n_j\in\mathcal N_j,n_q\in\mathcal N_q$, respectively, both of which are independent of the IRS reflection\footnote{Note that the single-reflection array response vector for path direction $(\theta,\phi)$ is comprised of the array response vector at the IRS and the channel coefficient for the link from the IRS to antenna array. While the double-reflection array response vector for path direction $(\theta,\phi)$ is comprised of the array response vector at the IRS, the channel coefficient for inter-IRS link, and that for the link from the IRS to antenna array.}. In practice, $\mv f_{j}^{n_{j}}(\theta,\phi)$ and $\mv g_{j,q}^{n_j,n_q}(\theta,\phi)$ can be obtained/estimated via various methods, such as the element-wise channel model \cite{my}, the ray-tracing technique \cite{ray_tracing}, or machine learning based on channel estimation method \cite{machine_learning}. In this paper, we adopt the element-wise channel proposed in \cite{my} to model $\mv f_{j}^{n_{j}}(\theta,\phi)$ and $\mv g_{j,q}^{n_j,n_q}(\theta,\phi)$, which are assumed to be known for the IRS reflection codebook design of our main interest in this paper.

Then, denote by $\vartheta_{j,n_j}$ the reflection coefficient of IRS$j$'s $n_j$-th reflecting element with $j\in\mathcal J, n_j\in\mathcal N_j$, where we set its amplitude to the maximum value of one, i.e., $|\vartheta_{j,n_j}|=1$, for the purpose of maximizing the reflected signal power, and $\mv\Theta\triangleq\{\vartheta_{j,n_j},~j\in\mathcal J,n_j\in\mathcal N_j\}$. Therefore, the  EARV at the IRS-AP for the path direction $(\theta,\phi)\in\mathcal A$ is given by \cite{my}
\begin{align}
\mv h(\theta,\phi,\mv\Theta)&=\mv h_{d}(\theta,\phi)+\sum_{j=1}^{J}\sum_{n_j=1}^{N_j}\mv f_{j}^{n_j}(\theta,\phi)\vartheta_{j,n_j}\nonumber\\
&+\sum_{j=1}^{J}\sum_{q\neq j}^{J}\sum_{n_j=1}^{N_j}\sum_{n_q=1}^{N_q}\mv g_{j,q}^{n_j,n_q}(\theta,\phi)\vartheta_{j,n_j}\vartheta_{q,n_q},\label{effective_gain}
\end{align}
which is related to IRS reflection coefficient, i.e., $\{\vartheta_{j,n_j}\}$.

\begin{table*}
\centering
\caption{List of Main Symbols and Their Physical Meanings}\label{notations}
\small
\begin{tabular}{|p{2cm}|p{5.5cm}||p{2cm}|p{5.5cm}|}
\hline
{\bf Symbol}&{\bf Physical meaning}&{\bf Symbol}&{\bf Physical meaning}\\
\hline
\hline
$M$&Dimension of AP antenna &$J$&IRS number\\
\hline
$N$&Dimension of IRS $j$&$d_{l}$/$d_{w}$/$d_t$&Length/width/thickness of the AP's antenna radome\\
\hline
$M_{x}$/$M_{y}$&Number of antennas along axis $x$ or axis $y$  &$N_{j,1}$/$N_{j,2}$&Number of reflecting elements at $j$-th IRS along axis $x$ (or $y$)/axis $z$\\
\hline
$H_{AR}$&Hight of the antenna radome from the ground&$A$&Reflecting element's aperture area\\
\hline
$\lambda$&Carrier wavelength&$d_{A}$/$d_{I}$&The spacing between two consecutive elements of the antenna array/IRSs\\
\hline
$G_{A}(\cdot)$&AP antennas' antenna gain&$\vartheta_{j,n_j}$&Reflection coefficient of IRS $j$'s $n_j$-th reflecting element\\
\hline
$\mathcal A$&Coverage area&$\mv h_{d}$&Direct array response vector\\
\hline
$\mv f_{j}^{n_{j}}$&Single-reflection array response vector via IRS $j$'s $n_j$-th reflecting element &$\mv g_{j,q}^{n_{j},n_{q}}$&Double-reflection array response vector via IRS $j$'s $n_j$-th reflecting element and then IRS $q$'s  $n_{q}$-th reflecting element \\
\hline
$\theta$/$\phi$&Elevation/azimuth AoA of a particular signal path from the coverage area $\mathcal A$ arriving at the IRS-AP&$\theta_{\psi}$/$\phi_{\psi}$&Elevation/azimuth angle of the $\psi$-th path at the IRS-AP\\
 \hline
 $a_{\psi}$&Complex coefficient of the $\psi$-th path&$\Psi$&Number of channel paths\\
\hline
$\tilde{\mv h}$&Effective channel&$\mv h_{LoS}(\cdot)$&Effective LoS channel\\
\hline
$\mv h(\cdot)$&Effective array response vector (EARV)&$D$&Codebook size for single-user transmission/Number of sectors\\
\hline
 $\mathcal{C}_{D}$&Codebook of size $D$ for single-user transmission&$\mathcal{A}_{D,d}$&$d$-th sector\\
 \hline
 $\mv\Theta^{*}_{D,d}$&$d$-th codeword in $\mathcal{C}_{D}$ for covering sector $\mathcal{A}_{D,d}$&$E_{D,d}(\mv\Theta_{D,d})$&SMAECP for sector $\mathcal{A}_{D,d}$\\
\hline
 $\tilde{\mathcal{C}}_{X}$&Codebook of size $X$ for multi-user transmission&$X$&Codebook size for multi-user transmission\\
 \hline
\end{tabular}
\end{table*}

Based on (\ref{effective_gain}), the effective channel between any location in the coverage area $\mathcal A$ and the IRS-AP can be modeled by the following multi-path channel, i.e.,
\begin{align}
\tilde{\mv h}=\sum_{\psi=1}^{\Psi}a_{\psi}\mv h(\theta_{\psi},\phi_{\psi},\mv\Theta), \label{effective_channel_old}
\end{align}
where $\Psi\geq 1$ denotes the total number of (significant) channel paths from any location in the coverage area $\mathcal A$ to the IRS-AP, $a_{\psi}$ represents the complex coefficient of the $\psi$-th path, $\theta_{\psi}\in[0,\theta_{\max}]$/$\phi_{\psi}\in[0,2\pi)$ denote the elevation/azimuth AoAs of the $\psi$-th path arriving at the IRS-AP, respectively. Specifically, we assume that the first path in (\ref{effective_channel_old}) corresponding to $\psi=1$ is the line-of-sight (LoS) path, while the remaining $\Psi-1$ paths are non-LoS (NLoS) paths. In particular, the complex coefficient of the LoS path between any location in the coverage area $\mathcal A$ and the AP is determined by their distance. As shown in Fig. \ref{system_model}, the horizontal coordinates for the location of the path direction $(\theta,\phi)\in\mathcal A$ can be expressed as $(H_{AR}\tan\theta\cos\phi,H_{AR}\tan\theta\sin\phi)$, and thus the distance between them is given by $\sqrt{(H_{AR}\tan\theta\cos\phi)^2+(H_{AR}\tan\theta\sin\phi)^2+H_{AR}^{2}}=H_{AR}/\cos\theta$, which is only dependent on the elevation angle $\theta$ of the path direction. As a result, the complex LoS path coefficient between the location of the path direction $(\theta,\phi)\in\mathcal A$ and the IRS-AP is given by 
\begin{align}
a_{1}=\frac{\lambda}{4\pi H_{AR}/\cos\theta}e^{-i\frac{2\pi}{\lambda}H_{AR}/\cos\theta},\label{a}
\end{align}
which is a function in terms of the elevation angle $\theta$. In contrast, the complex coefficients of other NLoS paths depend on the scatterers in the environment and will be modeled in Section \ref{simulation_results_section} for simulations. To facilitate comprehension, in Table \ref{notations}, we have compiled a summary of symbol notations utilized in this paper together with their corresponding physical meanings.

\vspace{-10pt}\section{Reflection Codebook Design Based on Sector Division}\label{codebook_design_section}
In this section, we propose a codebook-based IRS reflection design for the IRS-AP to enhance the average effective channel power of all locations in the coverage area $\mathcal A$. By assuming that the single-reflection and double-reflection array response vectors for all AoAs, i.e., $\{\mv f_{j}^{n_j}(\theta,\phi)\}$ and $\{\mv g_{j,q}^{n_j,n_q}(\theta,\phi)\}$, are available, a codebook consisting of a small number of IRS reflection patterns/codewords is first designed offline and stored at the IRS-AP. Then, when the users send orthogonal pilot signals to the IRS-AP for estimating their effective channels (i.e., $\tilde{\mv h}$ in (\ref{effective_channel_old})) as that in conventional cellular network without IRSs \cite{liuliang}, the IRS-AP can evaluate the communication performance based on the estimated channels by varying IRS reflection patterns according to the codebook\footnote{Notice that our proposed codebook-based reflection design only needs to estimate the effective channels between the antenna array of the AP and users under each codeword (i.e., IRS reflection pattern) and then selects the best one for data transmission. This is different from conventional channel estimation for IRS-assisted wireless communications \cite{channel_estimation0,channel_estimation2,liuliang,anchor,double_estimation,group,sparsity,channel_estimation_survey}, where the single-reflection and double-reflection channels should be separately estimated for IRS reflection design with efficient strategies, such as anchor-aided channel estimation, reference user-based channel estimation, channel estimation based on channel sparsity, and so on \cite{channel_estimation_survey}.}. As a result, the total channel estimation overhead is linear with the number of users and codebook size. Finally, the codeword achieving the best communication performance is selected for data transmission. To design the codebook efficiently, we only consider the LoS channel between all locations in the coverage area $\mathcal A$ and the IRS-AP. As a result, the effective LoS channel between the location of the path direction $(\theta,\phi)$ and the IRS-AP is given by
\begin{align}
&\mv h_{LoS}(\theta,\phi,\mv\Theta)=a_{1}\mv h(\theta_{1},\phi_{1},\mv\Theta)\nonumber\\
&=\frac{\lambda}{4\pi H_{AR}/\cos\theta}e^{-i\frac{2\pi}{\lambda}H_{AR}/\cos\theta}\mv h(\theta,\phi,\mv\Theta).
\end{align}
The reason for such consideration is that the LoS channel is deterministic and usually dominants in the channel power as compared to NLoS channels, while the NLoS channels depend on the scatterers in the propagation environment, which are in general randomly distributed and difficult to model \cite{nlos_no1,nlos_no2}. In the following, we first propose an efficient sector division strategy for achieving a small codebook size to reduce the channel training overhead, then explain the principle of codeword design for each divided sector, and finally present the procedure to construct the codebook for IRS reflection patterns.

Notice that in conventional reflection codebook design for IRS-assisted communication systems (see, e.g., \cite{codebook1,codebook3,codebook4,codebook5}), only the single-reflection paths via the IRSs exist between the users and AP.  Thus, the high-resolution and precise-alignment IRS reflection patterns/codewords can be designed to provide high directional gains to the single-reflection paths. Specifically,  the whole coverage area is divided into multiple non-overlapped sectors by equally dividing the intervals of the elevation and azimuth angles, which renders that each sector can be approximated by a single point in the angular domain. Then, an IRS reflection pattern/codeword is designed to align with the approximated point for each sector. Nevertheless, this sector division strategy will result in a large codebook size (e.g., hundreds to thousands) for IRS with similar number of reflecting elements, which needs extremely high overhead for IRS reflection training. In contrast, in our proposed IRS-AP, the double-reflection paths via different IRSs play a significant role due to the close distance among IRSs \cite{my}, which leads to no explicit directional gain obtained via IRS reflection design. As a result, it is practically viable to design the low-resolution and wide-coverage codewords for the IRS-AP, which entails a codebook with smaller size (i.e., only several to dozens) and thus greatly reduces the channel training overhead. 

Based on the above, we propose to divide the whole coverage area into multiple non-overlapped sector by {\it only} dividing the interval of the azimuth angle. Specifically, let $D$ denote the number of sectors for dividing the whole coverage area $\mathcal A$, where $\mathcal{D}=\{1,2,\cdots,D\}$ denotes the set of all sectors and $D$ is set as a small positive number to limit the codebook size. The division for the interval of the azimuth angle is given by
\begin{align}
[0,2\pi)&=\Big[0,\frac{2\pi}{D}\Big)\cup\cdots\cup\Big[(D-1)\times\frac{2\pi}{D},2\pi\Big).\label{divide}
\end{align} 
Then, the $d$-th sector can be expressed as
\begin{align}
&\mathcal A_{D,d}=\nonumber\\
&\Big\{(\theta,\phi)\Big|\theta\in\Big[0,\theta_{\max}\Big],\phi\in\Big[(d-1)\times\frac{2\pi}{D},d\times\frac{2\pi}{D}\Big)\Big\},\nonumber\\
&~d\in\mathcal D.\label{region_d}
\end{align}
Next, a codeword for IRS reflection pattern should be designed to ensure the coverage of sector $\mathcal A_{D,d}$. 
 
To this end, we propose to consider the worst-case performance in sector $\mathcal{A}_{D,d}$, which occurs at the locations with the largest elevation angle w.r.t. the antenna array, i.e., $\theta=\theta_{\max}$. This is because the path gain $|a_{1}|^2$ in (\ref{a}) decreases with the elevation angle $\theta$ due to the increasing distance $H_{AR}/\cos\theta$, which may cause severe near-far fairness issue among all elevation angles (i.e., a bottleneck problem for conventional multi-antenna AP without IRSs) if the IRS passive reflection is not properly set. Therefore, we focus on the performance of these locations at $\theta=\theta_{\max}$, which have the largest distance with the AP, i.e., $H_{AR}/\cos\theta_{\max}$, and thus the smallest path gain $|a_{1}|^{2}$ in (\ref{a}) with the AP. Moreover, we propose a performance metric for the maximum elevation angle $\theta_{\max}$, which is the average effective (LoS) channel power over all azimuth angles in sector $\mathcal{A}_{D,d}$ by fixing the elevation angle as $\theta_{\max}$, i.e.,
\begin{align}
E_{D,d}(\mv\Theta_{D,d})=\frac{D}{2\pi}\int_{\frac{2\pi}{D}(d-1)}^{\frac{2\pi}{D}d}||\mv h_{LoS}(\theta_{\max},\phi,\mv\Theta_{D,d})||^{2} d\phi,\label{average_channel_power}
\end{align}
where $\mv\Theta_{D,d}=\{\vartheta_{j,n_j}^{D,d},~j\in\mathcal J, n_j\in\mathcal N_{j}\}$ and $\vartheta_{j,n_j}^{D,d}$ denotes the corresponding reflection coefficient of  IRS $j$'s $n_j$-th reflecting element with its modulus as $1$, i.e., $|\vartheta_{j,n_j}^{D,d}|=1$. The reason for such consideration lies in that the (LoS) path gain $|a_{1}|^{2}$ in (\ref{a}) is the same over all azimuth angles $\phi$ under any given elevation angle $\theta$. For convenience, we define $E_{D,d}(\mv\Theta_{D,d})$ as {\it SMAECP} for sector $\mathcal A_{D,d}$, and consider the {\it SMAECP maximization} as the objective of designing the corresponding IRS reflection pattern/codeword. 

As such, the optimization problem for designing the IRS reflection pattern/codeword for sector $\mathcal{A}_{D,d}$, subject to the unit-modulus constraints for all IRS reflecting elements, can be formulated as follows.
\begin{align}
\text{(P1.$D$.$d$):}~\max_{\bm\Theta_{D,d}}&~E_{D,d}(\mv\Theta_{D,d})\nonumber\\
\text{s.t.}~&|\vartheta_{j,n_j}^{D,d}|=1,~j\in\mathcal J,~n_{j}\in\mathcal N_{j}.\label{unit}
\end{align}
Let $\mv\Theta_{D,d}^{*}$ denote the solution to problem (P1.$D$.$d$), with the details for obtaining it provided in Section \ref{codeword_design_section}.

In general, the effective (LoS) channel power of all paths from the coverage area $\mathcal A$, i.e., $||\mv h_{LoS}(\theta,\phi,\mv\Theta)||^{2},\theta\in[0,\theta_{\max}],\phi\in[0,2\pi)$, can be improved by applying the proposed codeword design obtained by solving (P1.$D$.$d$)'s. The performance improvement generally increases with the number of sectors $D$, which will be shown in Section \ref{simulation_results_section}.  This is because each codeword only needs to cover a smaller sector in the angular domain by increasing the number of sectors $D$, which helps improve the SMAECP for each sector, i.e., $E_{D,d}(\mv\Theta_{D,d}^{*}),~d\in\mathcal D$. 

\begin{figure}
\centering
\includegraphics[width=10cm]{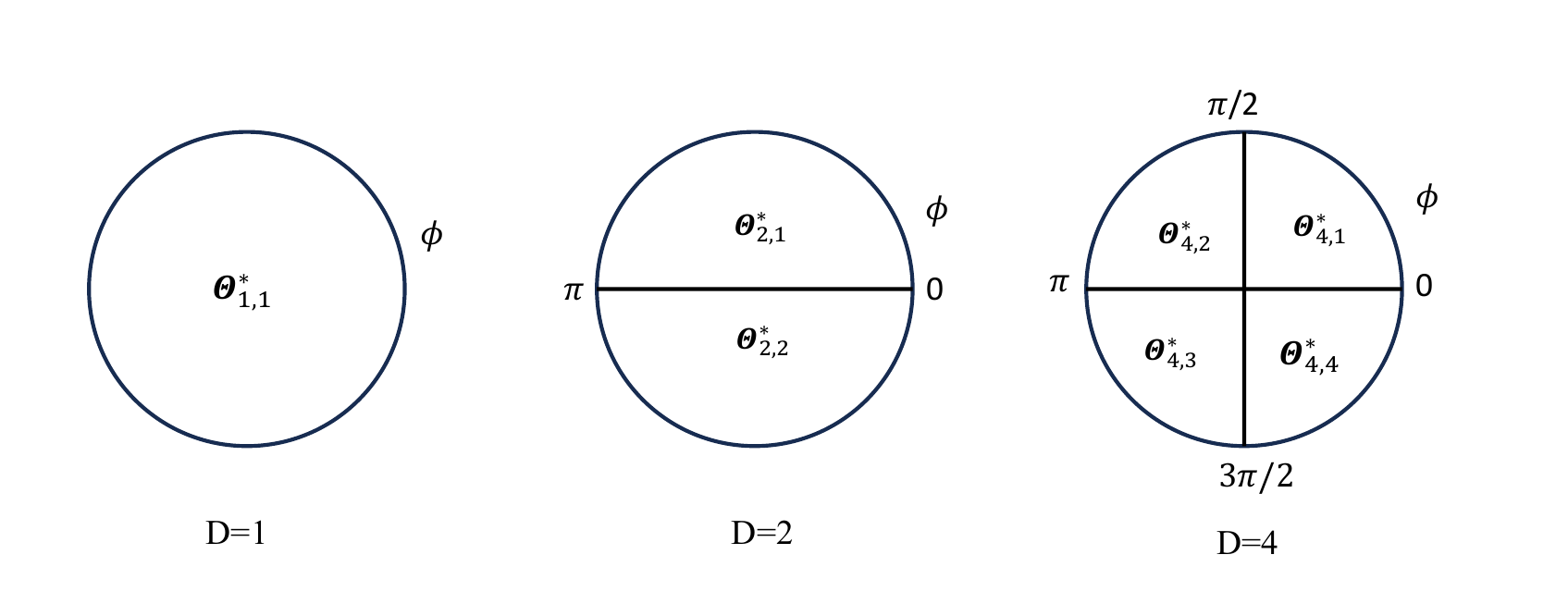}
\caption{Examples for codebook design based on sector division for single-user transmission (top-view).}\label{example_sector_division}
\end{figure}

As a result, for single-user transmission, the number of sectors $D$ can be considered as the codebook size, and the corresponding codebook can be constructed as $\mathcal{C}_{D}=\{\mv\Theta_{D,d}^{*},d\in\mathcal D\}$ with $\mv\Theta^{*}_{D,d}$ denoting its $d$-th codeword for covering sector $\mathcal{A}_{D,d}$, which is obtained via solving problem (P1.$D$.$d$). For ease of illustration, we show some examples in Fig. \ref{example_sector_division} by setting the codebook size (or equivalently the number of sectors) as $D=1$, $D=2$, and $D=4$, respectively. The codebook design for multi-user transmission based on the obtained codebook for single-user transmission, i.e., $\mathcal{C}_{D}$'s, will be specified later in Section \ref{simulation_results_section}.

\section{Proposed Solution to (P1.$D$.$d$)}\label{codeword_design_section}
In this section, we aim to solve the formulated problem (P1.$D$.$d$) for codeword design in Section \ref{codebook_design_section}. Notice that problem (P1.$D$.$d$) is difficult to be optimally solved due to the non-concave objective function and the non-convex unit-modulus constraints in (\ref{unit}). Besides, the objective function is a continuous integral function, which is hard to be dealt with. Furthermore, due to the existence of both single-reflection and double-reflection signals, the objective function intricately couple all IRSs reflecting elements' reflection coefficients. In the following, we first approximate the continuous integral in the objective function by a discrete-form summation, and then apply the AO and SDR methods to obtain an efficient solution to problem  (P1.$D$.$d$). Since all $D$ problems (P1.$D$.$d$) have a similar form and can be solved in parallel, we ignore the subscript $(\cdot)_{D,d}$ in $\mv\Theta_{D,d}$ and the superscript $(\cdot)^{D,d}$ in all $\vartheta_{j,n_j}^{D,d}$'s, and accordingly rewrite  (P1.$D$.$d$) as (P1) for notational simplicity.

To tackle the continuous integral in the objective function of (P1), we make an approximation for it by uniformly discretizing the interval of azimuth angle in $\mathcal{A}_{D,d}$, i.e., $[(d-1)\times\frac{2\pi}{D},d\times\frac{2\pi}{D})$, into $L$ subsets, where we denote $\mathcal L\triangleq\{1,2,\cdots,L\}$. As a result, problem (P1) can be approximated as 
\begin{align}
\text{(P2):}~\max_{\mv\Theta}~&~\frac{1}{L}\sum_{l=1}^{L}||\mv h_{LoS}(\theta_{\max},\phi_{l},\mv\Theta)||^{2}\nonumber\\
\text{s.t.}~&(\ref{unit}),\nonumber
\end{align}
where $\phi_{l}=(d-1)\times\frac{2\pi}{T}+(l-\frac{1}{2})\frac{2\pi}{DL}$. However, problem (P2) is still difficult to be solved since the reflection coefficients of all IRSs are coupled in the objective function. To tackle this challenge, we employ the AO method for iterative optimization of the reflection coefficients of a single IRS while keeping the reflection coefficients of the remaining $(J-1)$ IRSs fixed.

Under any given $\vartheta_{q,n_q},~q\neq j\in\mathcal J,n_q\in\mathcal N_q$, problem (P2) is reduced to the following optimization problem for designing the reflection coefficients of all reflecting elements at IRS $j$, i.e., 
\begin{align}
\text{(P2-$j$):}~\max_{\mv\vartheta_{j}}~&||\mv B_{j}\mv\vartheta_{j}+\mv c_{j}||^{2}\nonumber\\
\text{s.t.}~&|\vartheta_{j,n_j}|=1,~n_j\in\mathcal N_{j},
\end{align}
where $\mv\vartheta_{j}=[\vartheta_{j,1},\vartheta_{j,2},\cdots,\vartheta_{j,N_{j}}]^{T}$, $\mv B_{j}=[\mv b_{j,1},\mv b_{j,2},\cdots,\mv b_{j,N_j}]$ with 
\begin{align}
&\mv b_{j,n_j}=\frac{1}{L}\sum_{l=1}^{L}a(\theta_{\max})\Big (\mv f_{j}^{n_j}(\theta_{\max},\phi_{l})\nonumber\\
&+\sum_{q\neq j}\sum_{n_q=1}^{N_q}\mv g_{j,q}^{n_j,n_q}(\theta_{\max},\phi_{l})\vartheta_{q,n_q}\nonumber\\
&+\sum_{q\neq j}\sum_{n_q=1}^{N_q}\mv g_{q,j}^{n_q,n_j}(\theta_{\max},\phi_{l})\vartheta_{q,n_q}\Big),~n_j\in\mathcal N_{j},
\end{align}
and
\begin{align}
&\mv c_{j}=\frac{1}{L}\sum_{l=1}^{L}a(\theta_{\max})\Big(\mv h_d(\theta_{\max},\phi_{l})\nonumber\\
&+\sum_{q\neq j}\sum_{n_q=1}^{N_q}\mv f_{q}^{n_q}(\theta_{\max},\phi_{l})\vartheta_{q,n_q}\nonumber\\
&+\sum_{q\neq j}\sum_{r\neq j,r\neq q}\sum_{n_q=1}^{N_q}\sum_{n_r=1}^{N_r}\mv g_{q,r}^{n_q,n_r}(\theta_{\max},\phi_{l})\vartheta_{q,n_q}\vartheta_{r,n_r}\Big).
\end{align}
Notice that problem (P2-$j$) is still non-convex. To solve it, we adopt the SDR technique \cite{sdr}. Towards this end, we define 
\begin{align}\label{tildetheta}
\tilde{\mv B}_{j}=\Big[
\begin{matrix}
\mv B_{j}^{H}\mv B_{j}&\mv B_{j}^{H}\mv c_{j}\\
\mv c_{j}^{H}\mv B_{j}&0
\end{matrix}\Big],~\tilde{\mv\vartheta}_{j}=\Big[
\begin{matrix}
\mv\vartheta_{j}\\
1
\end{matrix}\Big].
\end{align}
Accordingly, (P2-$j$) can be equivalently transformed into 
\begin{subequations}
\begin{align}
\text{(P3-$j$):}~\max_{\tilde{\mv\vartheta}_{j}}~&\tilde{\mv\vartheta}_{j}^{H}\tilde{\mv B}_{j}\tilde{\mv\vartheta}_{j}\nonumber\\
\text{s.t.}~&|\tilde{\vartheta}_{j,n_j}|=1,~n_j\in\mathcal N_{j},\\
&\tilde{\vartheta}_{j,N_j+1}=1.
\end{align}
\end{subequations}
Furthermore, we define $\tilde{\mv\Theta}_{j}=\tilde{\mv\vartheta}_{j}\tilde{\mv\vartheta}_{j}^{H}$ with $\tilde{\mv\Theta}_{j}\succeq \mv 0$ and $\text{rank}(\tilde{\mv\Theta}_{j})=1$. As a result, problem (P3-$j$) can be further transformed into
\begin{subequations}
\begin{align}
\text{(P4-$j$):}~\max_{\tilde{\bm\Theta}_j}~&\text{trace}(\tilde{\mv B}_{j}\mv\Theta_{j})\nonumber\\
\text{s.t.}~&[\tilde{\mv \Theta}_{j}]_{n_j,n_j}=1,2,~n_j=1,\cdots,N_{j},N_{j}+1,\\
&\tilde{\mv\Theta}_{j}\succeq \mv 0,\\
&\text{rank}(\tilde{\mv\Theta}_{j})= 1.\label{rankone}
\end{align}
\end{subequations}
However, problem (P4-$j$) is still challenging to be optimally solved due to the non-convex rank-one constraint in (\ref{rankone}). To deal with it, we remove this constraint, and obtain a relax version of (P4-$j$), which is denoted as (P5-$j$) and can be optimally solved by existing convex optimization solvers such as CVX \cite{convex}. Denote by $\tilde{\mv\Theta}_{j}^{*}$ the optimal solution to (P5-$j$). 

Now, it remains to reconstruct the solution to problem (P4-$j$) or equivalently (P3-$j$) based on $\tilde{\mv\Theta}_{j}^{*}$. Specifically, if $\text{rank}(\tilde{\mv\Theta}_{j}^{*})=1$, then $\tilde{\mv\Theta}_{j}^{*}$ serves as the optimal solution to (P4-$j$). In this case, we have $\tilde{\mv\Theta}_{j}^{*}=\tilde{\mv\vartheta}_{j}^{*}\tilde{\mv\vartheta}_{j}^{*}$, where $\tilde{\mv\vartheta}_{j}^{*}$ becomes the optimal solution to (P3-$j$). However, if $\text{rank}(\tilde{\mv\Theta}_{j}^{*})>1$, we need to employ the Gaussian randomization procedure to reconstruct a high-quality rank-one solution to problem (P4-$j$) or (P3-$j$). Specifically, assuming the eigenvalue decomposition of $\tilde{\mv\Theta}_{j}^{*}$ as $\tilde{\mv\Theta}_{j}^{*}=\mv U\mv \Sigma\mv U^{H}$, we set $\hat{\mv\vartheta}_{j}=\mv U\mv \Sigma^{\frac{1}{2}}\mv r$ with $\mv r\sim\mathcal{CN}(0,\mv I_{N_{j+1}})$. Therefore, we construct a feasible solution to (P3-$j$) as $\tilde{\vartheta}_{j,n_j}=e^{i\text{arg}(\hat{\vartheta}_{j,n_j}/\hat{\vartheta}_{j,N_{j}+1})},~n_{j}=1,2,\cdots,N_{j}+1$. To ensure the performance, the randomization process needs to be repeated numerous times and the best-selected solution to problem (P3-$j$) is denoted as $\mv{\tilde{\vartheta}}_{j}^{*}$. Using the solution of $\mv{\tilde{\vartheta}}_{j}^{*}$ to problem (P3-$j$), we can accordingly obtain the solution to (P2-$j$) as $\mv{\vartheta}_{j}^{*}$ based on (\ref{tildetheta}). Consequently, the SDR-based algorithm for solving problem (P2-$j$) is completed.

After obtaining the solution to (P2-$j$) as $\mv{\vartheta}_{j}^{*}$, we proceed to derive our proposed solution to (P2). To begin, we generate  $\Gamma$ solutions for $\mv\Theta$ to satisfy $|\vartheta_{j,n_j}|=1,~j\in\mathcal J,n_j\in\mathcal N_{j}$, where each $\vartheta_{j,n_j}$'s phase shift follows a uniform distribution in the interval $[0,2\pi)$. Next, we choose the solution, which achieves the largest objective value of (P2), as our initial point. Then, we sequentially solve problem (P2-$j$) for $j$ ranging from 1 to $J$ to update IRS reflecting elements' reflection coefficients. Once the reflection coefficients of all IRSs' reflecting elements  have been updated, we evaluate whether the increase in the objective value of (P2) is less than a specific small positive threshold $\epsilon$, or if the iteration number has reached the maximum allowable number of iteration $I_{\max}$. If either condition is met, the algorithm outputs the solution to (P2), denoted as $\mv\Theta^{*}_{D,d}$; otherwise, we repeat the above process. The overall algorithm is summarized in Algorithm \ref{ao}. It can be shown that the complexity of solving problem (P2-$j$) via the SDR method is given by $\mathcal{O}(N_{j}^{2}M+N_{j}^{3})$ \cite{complex}, and thus the overall complexity\footnote{Notice that the codeword design can be implemented offline, which does not add the time complexity to real-time implementation of effective channel estimation and codeword selection.} of Algorithm \ref{ao} to solve (P2) (and accordingly (P1)) is given by $\mathcal{O}(I\sum_{j=1}^{J}(N_{j}^{2}M+N_{j}^{3}))$, where $I\leq I_{\max}$ denotes the number of iterations required to achieve the convergence of Algorithm \ref{ao}.
\begin{algorithm}
\caption{Proposed Solution to (P1.$D$.$d$).}
\label{ao}
\begin{algorithmic}[1]
\REQUIRE {$a_{1}$, $\mv h_{d}(\theta_{\max},\phi)$, $\mv f_{j}^{n_j}(\theta_{\max},\phi)$, $\mv g_{j,q}^{n_j,n_q}(\theta_{\max},\phi)$, $\Gamma$, and $L$.}
\STATE{Obtain the sampling points of azimuth angles as $\{\phi_{l}\}$, and accordingly transform problem (P1) to (P2).}
\STATE{Randomly generate $\Gamma$ independent realizations of $\mv\Theta$ and obtain the corresponding objective value of (P2).}

\STATE{Select $\mv\Theta^{\star}$ as the realization yielding the largest objective value of (P2).}
\STATE{Initialize $\mv\Theta=\mv\Theta^{\star}$.}
\REPEAT
\FOR{$j=1\rightarrow J$}
\STATE{Obtain the solution to (P2-$j$), i.e., $\mv\vartheta_{j}^{*}$, by applying SDR algorithm.}
\STATE{Update $\mv\vartheta_{j}=\mv\vartheta_{j}^{*}$. }
\ENDFOR
\UNTIL{the fractional increase of the objective function in (P2) is below a threshold $\epsilon$ or the iteration number reaches the maximum allowable number of iterations $I_{\max}$. }
\STATE{Update $\mv\Theta^{*}_{D,d}=\mv\Theta$}
\ENSURE {$\mv\Theta^{*}_{D,d}$.}
\end{algorithmic}
\end{algorithm}

\section{Numerical Results}\label{simulation_results_section}
In this section, we provide numerical results to validate the performance of our proposed codebook-based IRS reflection design for the IRS-AP. Unless otherwise specified, we set the simulation parameters as follows. We set the carrier frequency as $f_c=6~\text{GHz}$, which leads to the corresponding wavelength as $\lambda=0.05~\text{m}$. The antenna radome's hight from the ground is set as $H_{AR}=5~\text{m}$. The dimension of the AP's antenna array is set as $M=4$ with $M_{x}=M_{y}=2$. The spacing between two adjacent AP's antennas is established as $d_{A}=\lambda/2=0.025~\text{m}$. The length, width, and thickness of the antenna radome are respectively set as $d_{l}=d_{w}=5\lambda$ and $d_{t}=\lambda/2$. The maximum elevation angle of all paths from the coverage area $\mathcal A$ is set as $\theta_{\max}=4\pi/9$. The element spacing within each IRS is set as $d_{I}=\lambda/2=0.025~\text{m}$, and each reflecting element's aperture area is set as $A=(\lambda/2)^2$. According to Remark \ref{deployment}, we set $N_{j,1}=10$ and $N_{j,2}=1$, and thus $N_{j}=10$ for each IRS $j,~j\in\mathcal J$. In the context of solving problem (P1.$D$.$d$), we set $L=40$. We set the stopping threshold and maximum number of iterations for Algorithm \ref{ao} as $\epsilon=10^{-5}$  and $I_{\max}=100$, respectively. The number of random initialization is set as $\Gamma=100$. Furthermore, the half-isotropic radiation pattern is adopted for AP's antennas, for which we have $G_{A}(\theta,\phi)=2$  for $0\leq \theta\leq \pi/2$ and $G_{A}(\theta,\phi)=0$ otherwise. Besides, we set the channel coherence time to be sufficiently large so that the effect of the channel estimation overhead on user rate performance is marginal. All results are computed as average across 100 random and independent user distributions and channel realizations.
 
\subsection{Performance Evaluation of Codebook Design}
\begin{figure}
\centering
\includegraphics[width=7cm]{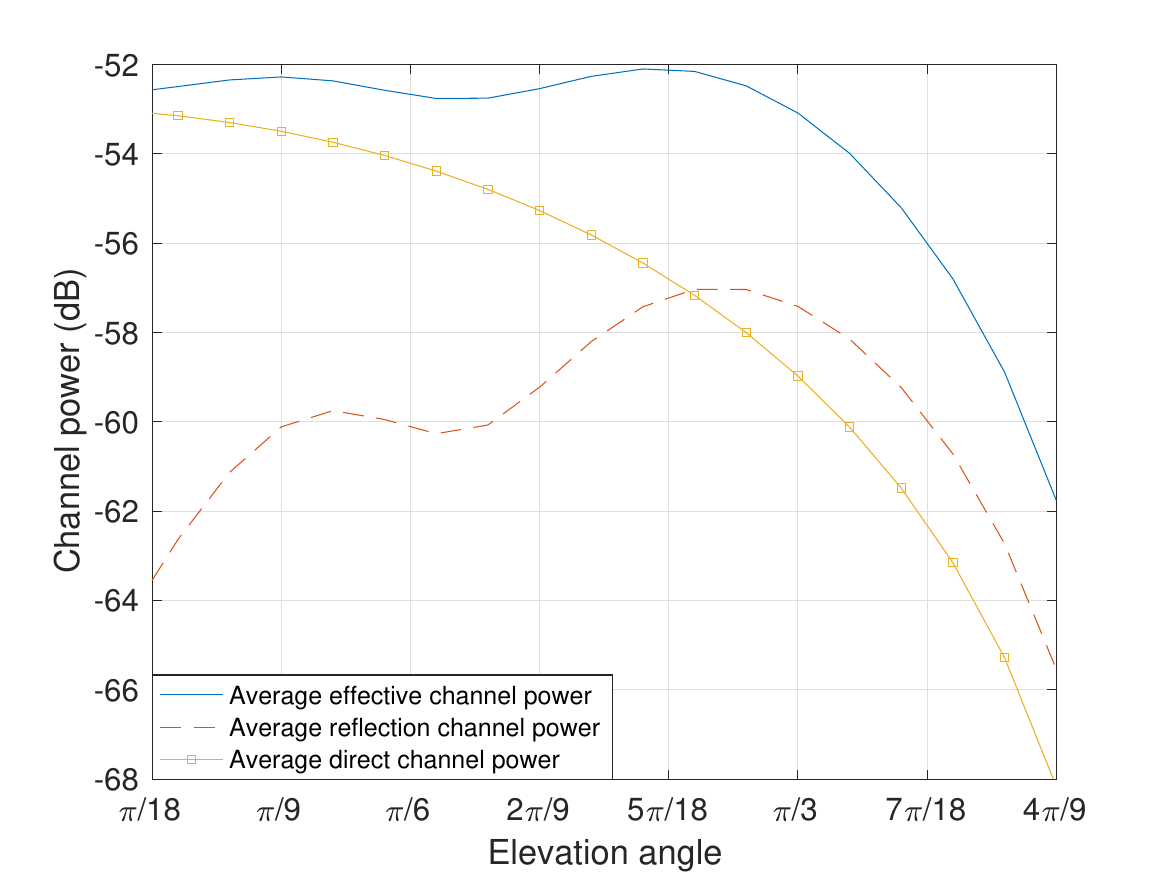}
\caption{Elevation power pattern of the designed codeword $\mv \Theta_{4,1}^{*}$.}\label{elevation_pattern}
\end{figure}

In this subsection, we aim to evaluate the performance of our proposed codebook design based on sector division in Section \ref{codebook_design_section}. First, we plot the elevation power pattern of the designed codeword $\mv\Theta_{4,1}^{*}$ in Fig. \ref{elevation_pattern}, which is the codeword for sector $\mathcal A_{4,1}$ by setting the number of sectors as $D=4$. To better demonstrate the effectiveness of our proposed codeword design via solving (P1.$D$.$d$), we consider three different channel power for each elevation angle $\theta$ as follows.
\begin{itemize}
\item{\bf Average effective channel power:} $\frac{2}{\pi}\int_{0}^{\pi/2}|a_{1}|^2||\mv h(\theta,\phi,\mv\Theta_{4,1}^{*})||^{2} d\phi$.
\item{\bf Average reflection channel power:} $\frac{2}{\pi}\int_{0}^{\pi/2}|a_{1}|^{2}||\mv h(\theta,\phi,\mv\Theta_{4,1}^{*})-\mv h_{d}(\theta,\phi)||^{2} d\phi$.
\item{\bf  Average direct channel power:} $\frac{2}{\pi}\int_{0}^{\pi/2}|a_{1}|^{2}||\mv h_{d}(\theta,\phi)||^{2} d\phi$.
\end{itemize} 
It is observed from Fig. \ref{elevation_pattern} that the average direct channel power decreases with the elevation angle $\theta$, since the path gain, i.e., $|a_{1}|^2$ in (\ref{a}), decreases with $\theta$ as the distance $H_{AR}/\cos\theta$ becomes longer. This demonstrates the severe near-far propagation issue in conventional multi-antenna AP systems without IRSs, and also validates the rationality of considering worst-case performance improvement as the objective for codeword design (see problem (P1.$D$.$d$)). Besides, it is observed that the average reflection channel power is small when $\theta\leq \pi/6$ due to the weaker IRS reflection gain for these locations. However, these locations do not need the reflection gain from the IRSs as compared to farther locations since they have strong enough average direct channel power with the AP owing to the shorter distance between them. In addition, the reflection channel power is observed to be limited when $\theta\geq 7\pi/18$ due to the longer distance with the AP. Thus, the locations with larger elevation angles w.r.t. the AP will also limit the average effective channel power. Moreover, the average effective channel power is observed to be larger than the average direct channel power under all considered $\theta$'s, even though the proposed codeword design only focuses on improving the worst-case  performance (i.e., at the largest elevation angle) in each sector (see problem (P1.$D$.$d$)). Furthermore, the worst-case performance at the elevation angle with $\theta=\theta_{\max}$ is observed to be improved by $6~\text{dB}$ based on the designed codeword $\mv\Theta_{4,1}^{*}$.

\begin{figure}
\centering
\includegraphics[width=7cm]{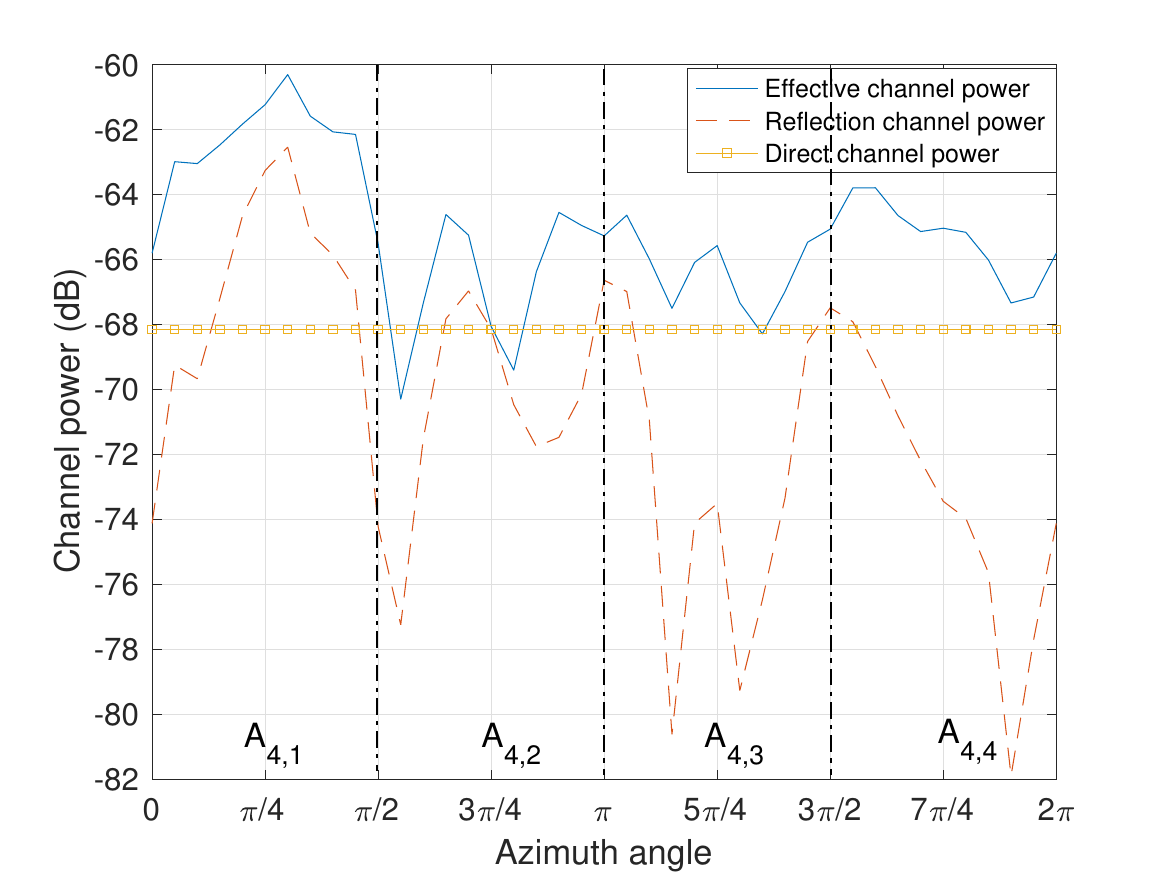}
\caption{Azimuth power pattern of the designed codeword $\mv \Theta_{4,1}^{*}$.}\label{azimuth_pattern}
\end{figure}

Next, Fig. \ref{azimuth_pattern} shows the azimuth power pattern of the designed codeword $\mv\Theta_{4,1}^{*}$ by fixing the elevation angle as $\theta=\theta_{\max}$, where we consider the following three different channel power for each azimuth angle $\phi$.
\begin{itemize}
\item {\bf Effective channel power:} $|a_{1}|^{2}||\mv h(\theta_{\max},\phi,\mv\Theta_{4,1}^{*})||^{2}$.
\item {\bf Reflection channel power:} $|a_{1}|^{2}||\mv h(\theta_{\max},\phi,\mv\Theta_{4,1}^{*})-\mv h_{d}(\theta_{\max},\phi)||^{2}$.
\item {\bf Direct channel power:} $|a_{1}|^{2}||\mv h_{d}(\theta_{\max},\phi)||^{2}$.
\end{itemize}
It is observed from Fig. \ref{azimuth_pattern} that the direct channel power is the same for all azimuth angles since they have the same distance with the AP, i.e., $H_{AR}/\cos\theta_{\max}$, which leads to the same path gain with the AP, i.e., $|a_{1}|^{2}$ (see (\ref{a})). Besides, it is observed that the effective channel power is much higher in sector $\mathcal {A}_{4,1}$, i.e., $0\leq \phi< \pi/2$, as compared to other sectors, i.e., $\mathcal{A}_{4,2}$ with $\pi/2\leq \phi<\pi$, $\mathcal{A}_{4,3}$ with $\pi\leq \phi<3\pi/2$, and $\mathcal{A}_{4,4}$ with $3\pi/2\leq \phi<2\pi$. This is because the designed codeword $\mv\Theta_{4,1}^{*}$ focuses on enhancing the channel power of all locations in sector $\mathcal{A}_{4,1}$ via solving problem (P1.4.1). In this case, the reflection channel power for all locations in sector $\mathcal{A}_{4,1}$ can be more significantly enhanced with the IRS reflection pattern as $\mv\Theta_{4,1}^{*}$. Besides, the single-reflection and double-reflection signals from all locations in $\mathcal{A}_{4,1}$ via the IRSs with the pattern as $\mv\Theta_{4,1}^{*}$ are almost in-phase with those propagated through their direct channels, and thus the effective channel power of all locations in $\mathcal{A}_{4,1}$ can be notably improved. However, for the locations in sectors $\mathcal{A}_{4,2}$, $\mathcal{A}_{4,3}$, and $\mathcal{A}_{4,4}$, their reflection channel power achieved by IRS reflection pattern as $\mv\Theta_{4,1}^{*}$ is much lower, and the reflection signals via the IRSs with the pattern as $\mv\Theta_{4,1}^{*}$ can be combined either constructively or destructively with that via direct channel, both of which result in less enhancement or even loss in their effective channel power as compared to direct channel power. 

Then, for performance comparison, we consider the following benchmark schemes.
\begin{itemize}
\item {\bf Random codebook:} In this benchmark scheme, the codebook size is set as $D$, and all codewords are randomly generated to satisfy the unit-modulus constraint, where the phase shifts of all reflecting elements in each codeword follow the uniform distribution in the interval $[0,2\pi)$. Similarly, the codeword with the best performance is employed for IRS reflection during user d ta transmission.
\item {\bf Discrete Fourier transform (DFT) codebook:} In this benchmark algorithm, we search the optimal codeword in a two-dimensional DFT codebook for designing the IRS passive reflection coefficients. For each IRS $j$, the codebook is defined as $\mathcal W_{j}=\{\mv w_{j}|\mv w_{j}=\mv w_{j,1}\otimes\mv w_{j,2},\mv w_{j,1}\in\mathcal{W}_{j,1},\mv w_{j,2}\in\mathcal{W}_{j,2}\}$, where $\mathcal{W}_{j,1}$ and $\mathcal{W}_{j,2}$ represent the sets of vectors including all the columns of DFT matrices of size $N_{j,1}$ and $N_{j,2}$, respectively. Then, we jointly search the passive reflection coefficients of all IRSs over their given codebooks for achieving the best communication performance. Thus, the overall training overhead required for implementing DFT codebook is given by $\prod_{j=1}^{J}N_{j,1}N_{j,2}$.
\item {\bf Unity reflection:} In this benchmark scheme, we set the reflection coefficients of all reflecting elements as 1, i.e., $\vartheta_{j,n_j}=1,~j\in\mathcal J,n_j\in\mathcal N_{j}$.
\item {\bf No-IRS:} In this benchmark scheme, we consider the conventional multi-antenna AP system without integrated IRSs.
\end{itemize}

\begin{figure}
\centering
\includegraphics[width=7cm]{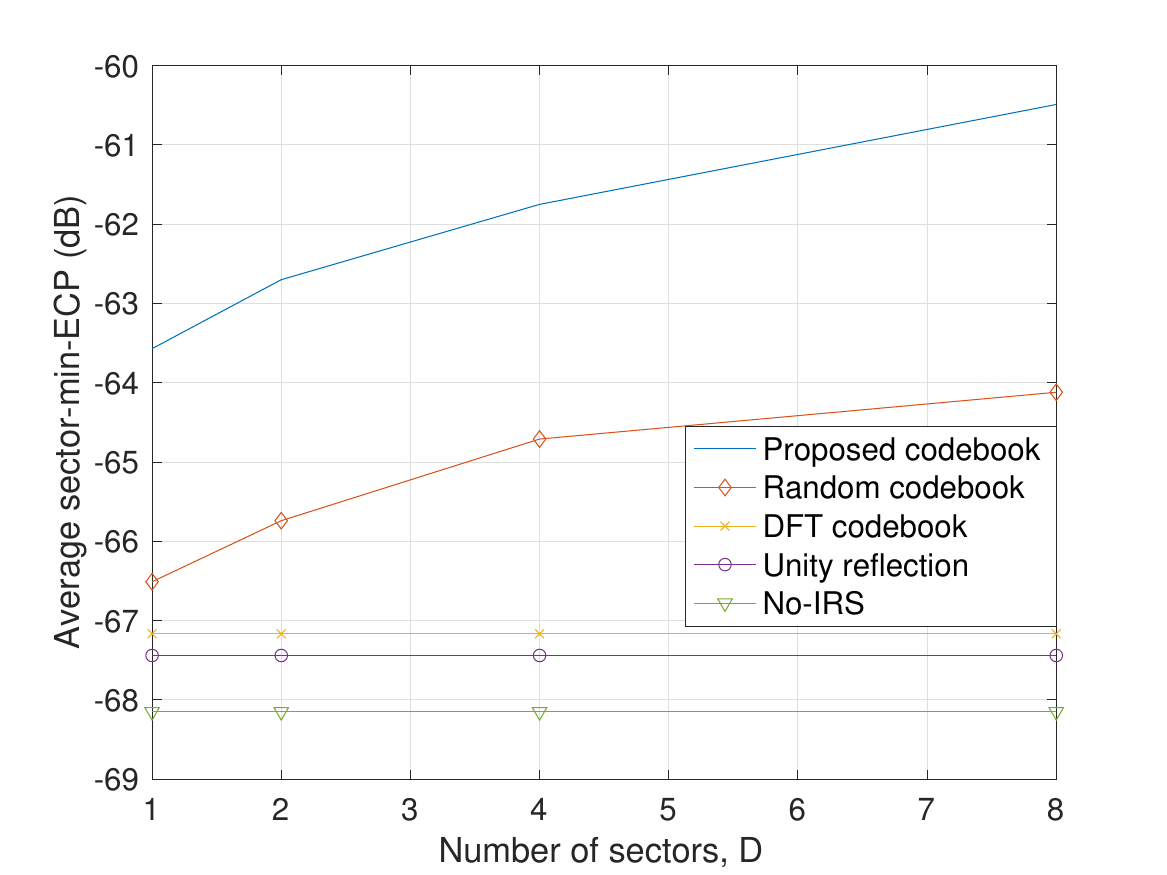}
\caption{Average SMAECP versus the number of sectors $D$.}\label{antenna_power_size_codebook}
\end{figure} 

Fig. \ref{antenna_power_size_codebook} shows the average SMAECP, i.e., $\frac{1}{D}\sum_{d=1}^{D}E_{D,d}(\mv\Theta_{D,d}^{*})$, of the whole coverage area $\mathcal A$ achieved by all considered schemes versus the number of sectors $D$. It is observed that our proposed codebook design outperforms all benchmark schemes under all considered $D$'s. Specifically, when the number of sectors is set as $D=8$, our proposed codebook design can achieve $3.63~\text{dB}$, $6.67~\text{dB}$, $6.95~\text{dB}$, and $7.66~\text{dB}$ gain over random codebook, DFT codebook, unity reflection, and no-IRS schemes, respectively. It is also observed that the average SMAECP can be improved by increasing the number of sectors $D$ in our proposed codebook design. Since each codeword only needs to cover a smaller sector by increasing the number of sectors $D$, which increases its alignment accuracy and thus better enhances the SMAECP. Furthermore, all schemes with IRSs are observed to outperform the no-IRS scheme, which indicates the IRS-AP is an efficient solution to improve the coverage performance.

\subsection{Performance Evaluation for Single-User Transmission}
In this subsection, we aim to evaluate the performance of our proposed codebook-based IRS reflection design for single-user transmission. For performance comparison, the upper-bound performance is considered in the following simulations, where the AoAs and complex coefficients of all paths are assumed to be perfectly estimated at the AP for constructing the effective channel with the user, and the successive refinement algorithm proposed in \cite{my} is adopted for IRS reflection design, termed as ``perfect CSI''. We adopt the Rician fading channel to generate the effective channel with the user in (\ref{effective_channel_old}) by setting $\sum_{\psi=2}^{\Psi}\mathbb{E}(|a_{\psi}|^{2})/|a_{1}|^{2}=1/\kappa$ and $a_{\psi}=\frac{|a_{1}|}{\sqrt{\kappa(\Psi-1)}}\mathcal{CN}(0,1)$ with $\psi=2,3,\cdots,\Psi$, where $\kappa\geq 0$ denotes the Rician factor. Besides, we fix $\theta=\theta_{\max}$ (i.e., worst-case user locations) and randomly generate $\phi$ from interval $[0,2\pi)$ to consider the cell-edge user. In addition, we randomly generate $\theta_{\psi}$ and $\phi_{\psi},~\psi=2,\cdots,\Psi$, from intervals $[0,\theta_{\max}]$ and $[0,2\pi)$, respectively, to account for random scattering environment, where the total number of signal propagation paths is set as $\Psi=5$. Moreover, the performance metric is selected as the achievable rate of the user, i.e., $R=\log_{2}\Big(1+\frac{P||\tilde{\bm h}||^{2}}{\sigma^{2}}\Big)$, where $P$ denotes the transmit power of the user and $\sigma^{2}$ denotes the noise power at the AP's antenna array. 

\subsubsection{Effect of Channel Estimation Overhead}
\begin{figure}
\centering
\includegraphics[width=7cm]{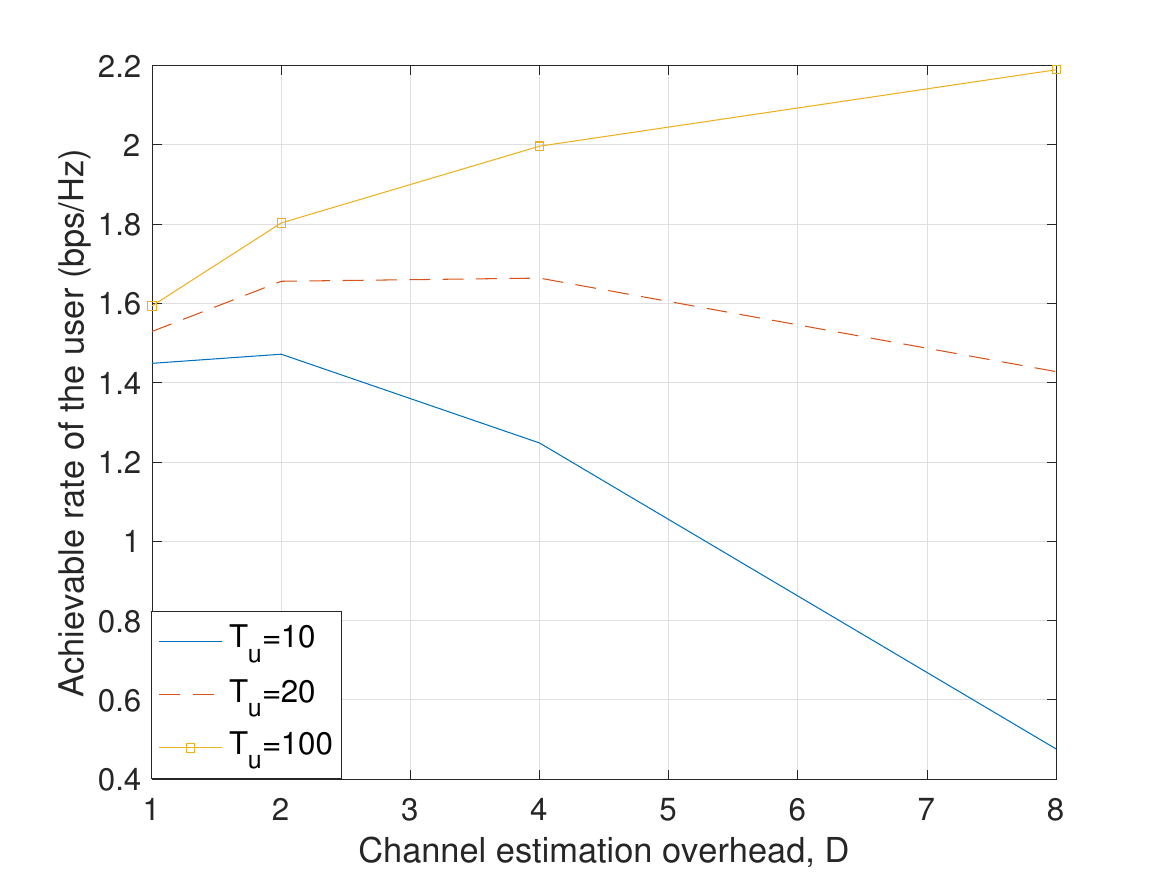}
\caption{Effect of channel estimation overhead.}\label{SIMO_D_T}
\end{figure}

First, we aim to evaluate the effect of channel estimation overhead on user's achievable rate. Since only effective channel with the user is estimated for each codeword, the channel estimation overhead is given by $D$ in single-user transmission. Fig. \ref{SIMO_D_T} shows the achievable rate of the user achieved by our proposed codebook-based reflection design versus the channel estimation overhead (i.e., the codebook size) $D$ under different channel coherence time, denoted as $T_{u}$, where the Rician factor is set as $\kappa=10~\text{dB}$. It is observed that when $T_{u}$ is sufficiently small, i.e., $T_{u}=10$, the user's achievable rate decreases with increasing $D$ for our proposed codebook-based reflection design due to the more dominant effect of the data transmission time on the user's achievable rate than user's signal-to-noise-ratio (SNR). However, when $T_{u}$ becomes large, i.e., $T_{u}=100$, it is observed that the user's achievable rate monotonically increases with $D$. This is because in this case, the channel estimation overhead, i.e., $D$, is much smaller than $T_{u}$. As a result, the user's achievable rate is more dominated by its SNR, which can be improved as $D$ increasing.  Furthermore, when $T_{u}=20$, it is observed that the achievable rate of the user first increases with $D$ when $D\leq 4$ and then decreases with it. This indicates that in the case of a moderate $T_{u}$, the data transmission time and user SNR should be optimally balanced to maximize the user achievable rate.

\subsubsection{Effect of Codebook Size}

\begin{figure}
\centering
\includegraphics[width=7cm]{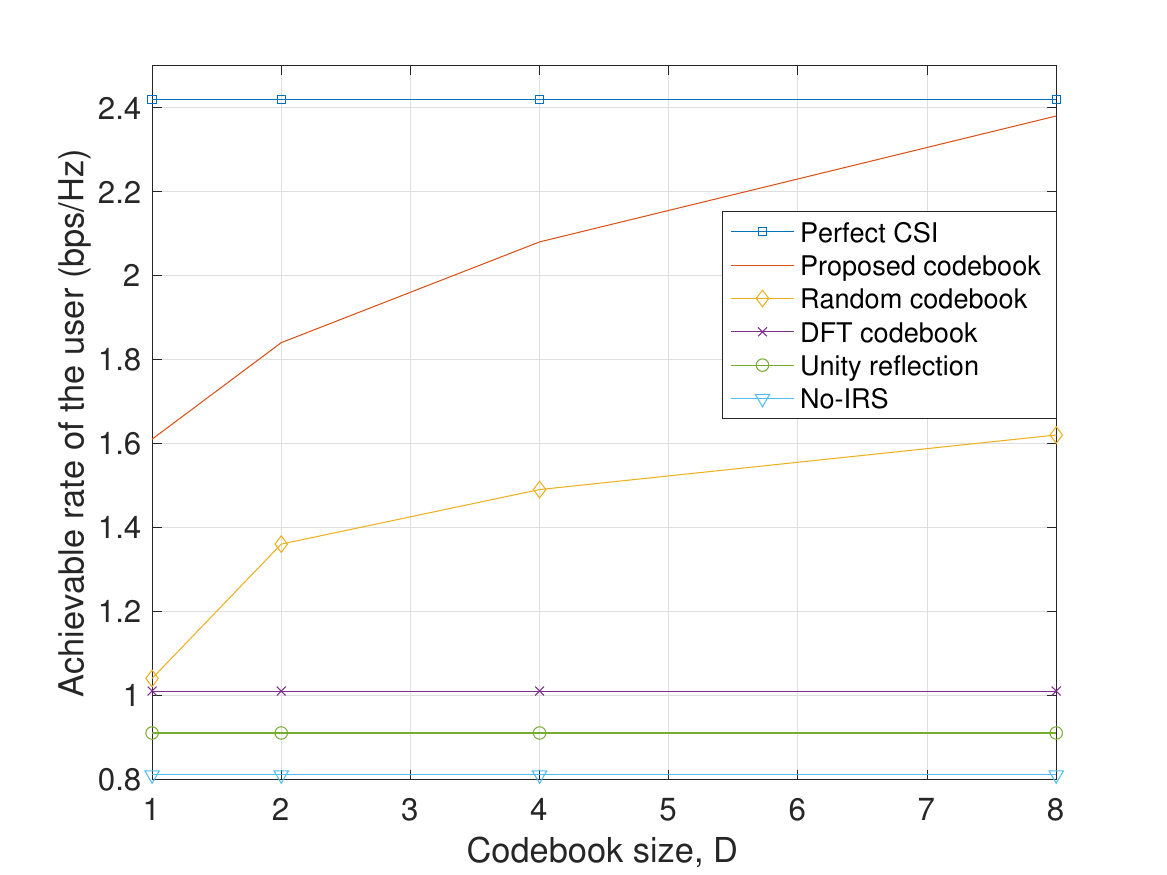}
\caption{Achievable rate of the user versus codebook size $D$.}\label{SIMO_D}
\end{figure}

Next, Fig. \ref{SIMO_D} shows the achievable rate of the user versus the codebook size $D$ for all considered schemes, where the Rician factor is set as $\kappa=10~\text{dB}$. As we can see, the achievable rate of the user increases with $D$ in our proposed codebook-based reflection design, since the codewords achieving higher SMAECP can be generated to better enhance the effective channel power of the user, which leads to a higher achievable rate. The achievable rate of the user for the random codebook-based reflection design is also observed to increase with $D$. This is because more codeword candidates are randomly generated, which enables a higher chance to select a codeword to achieve better performance. In addition, our proposed codebook-based reflection design is observed to outperform all benchmark schemes under all considered $D$'s. Specifically, when $D=8$, our proposed codebook-based reflection design has 46.91\%, 135.63\%, 161.51\%, and 193.83\% rate improvements over random codebook, DFT codebook, unity reflection, and no-IRS schemes, respectively. Notice that the conventional DFT codebook cannot achieve better performance in our proposed IRS-AP since the far-field channel condition is not satisfied between the IRSs and antenna array of the AP due to the ultra-short distance between them, and the more dominant double-reflection channels coexist with the single-reflection channels. Furthermore, it is observed that the achievable rate in our proposed codebook-based reflection design can approach that in the scheme with perfect CSI by setting the codebook size as $D=8$, which thus greatly reduces the channel training overhead required for obtaining perfect CSI.

\subsubsection{Effect of Rician Factor}
\begin{figure}
\centering
\includegraphics[width=7cm]{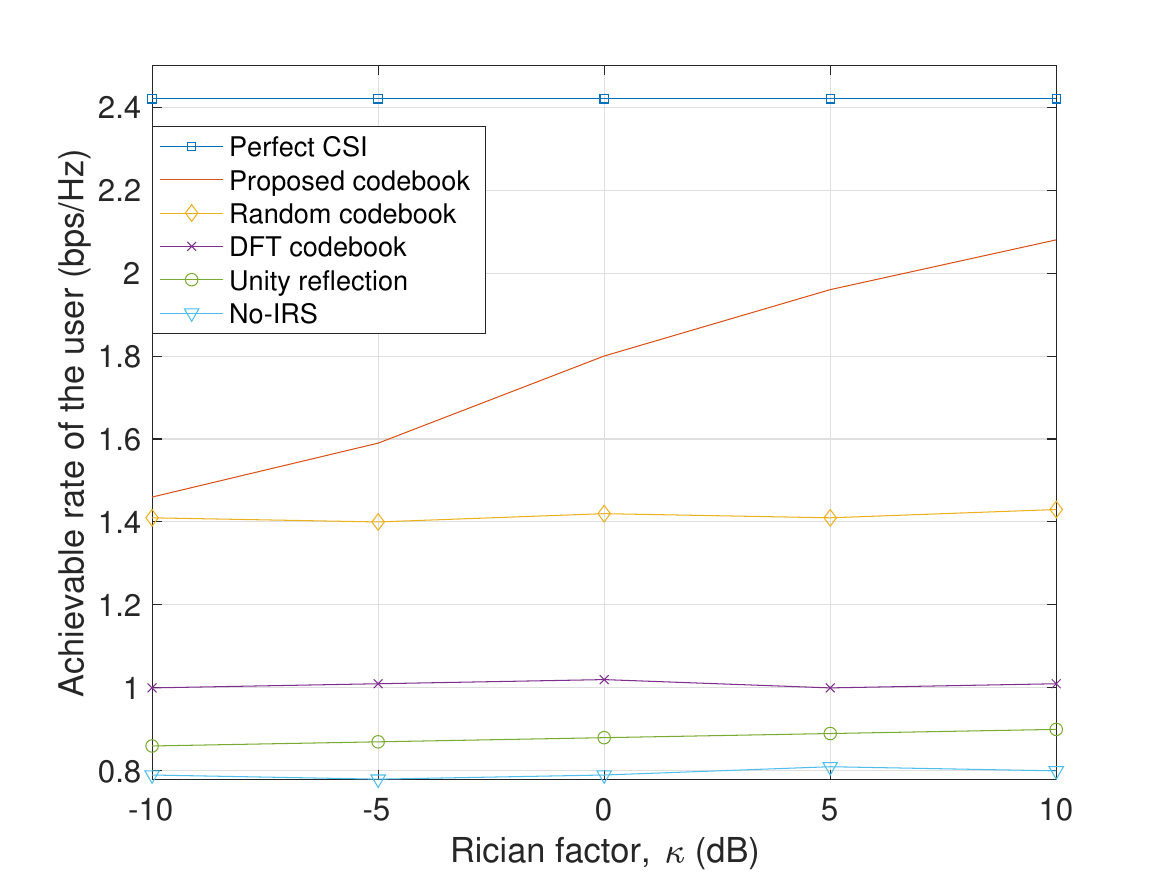}
\caption{Achievable rate of the user versus Rician factor $\kappa$.}\label{SIMO_K}
\end{figure}

Then, Fig. \ref{SIMO_K} shows the effect of Rician factor $\kappa$ on the achievable rate of the user for all schemes, where we set the codebook size as $D=4$. It is observed that the performance achieved by our proposed codebook-based reflection design increases with Rician factor $\kappa$ due to the stronger LoS channel. In addition, our proposed codebook-based reflection design is observed to achieve a better performance than other benchmark schemes under all considered $\kappa$'s, even for strong NLoS environment (i.e., $\kappa=0~\text{dB}$). This is because our proposed codebook-based reflection design can enhance the channel power of all paths from the coverage area $\mathcal A$, including the LoS channel and all NLoS channels (in general, see Fig. \ref{azimuth_pattern}), and thus the achievable rate of the user in our proposed codebook-based reflection design can be better enhanced.

\subsubsection{Slow Adaptation for IRS Passive Reflection}
Furthermore, we aim to show that our proposed codebook-based IRS reflection design can achieve considerable communication performance improvement even with slow adaptation to wireless channels. Specifically, by assuming that all involved channels are constant during each fading block, the AP only needs to implement the IRS reflection training at the first fading block, then the IRS reflection pattern can remain unchanged for a long time (e.g., in the order of hundreds to thousands of fading blocks), which can help reduce the channel training overhead and the controlling overhead for adjusting IRS reflection pattern. The reason for such consideration lies in that our proposed codebook design is based on sector division (see Section \ref{codebook_design_section}), where the codeword is generated to ensure the performance of all locations within its corresponding sector (see Problem (P1.$D$.$d$)). As a result, as long as the user stays in the same sector, the IRS reflection pattern does not need to be changed for ensuring the average performance of the user. 

Towards this end, we consider a scenario with a low-mobility user by fixing its elevation angle as $\theta=\theta_{\max}$ and changing its azimuth angle from $0$ to $2\pi$ with the angular speed of $\frac{\pi}{36}$ rad/s. As a result, the linear speed of the user is given by $v=2~\text{m/s}$, which results in a Doppler frequency with the maximum value of $f_{\max}=vf_{c}/c= 40 ~\text{Hz}$, where $c=3\times 10^{8}~\text{m/s}$ denotes the speed of light. The duration of each fading block is set as $1/(10f_{\max})=0.0025~\text{s}$, and thus each instant includes 400 fading block. Let $T>0$ denote the time duration of fixing an IRS reflection pattern, i.e., the IRS reflection is updated once after $T~\text{s}$. 
\begin{figure}
	\centering
	\subfigure[$D=8$]{
		\begin{minipage}[t]{0.45\textwidth}
			\includegraphics[width=7cm]{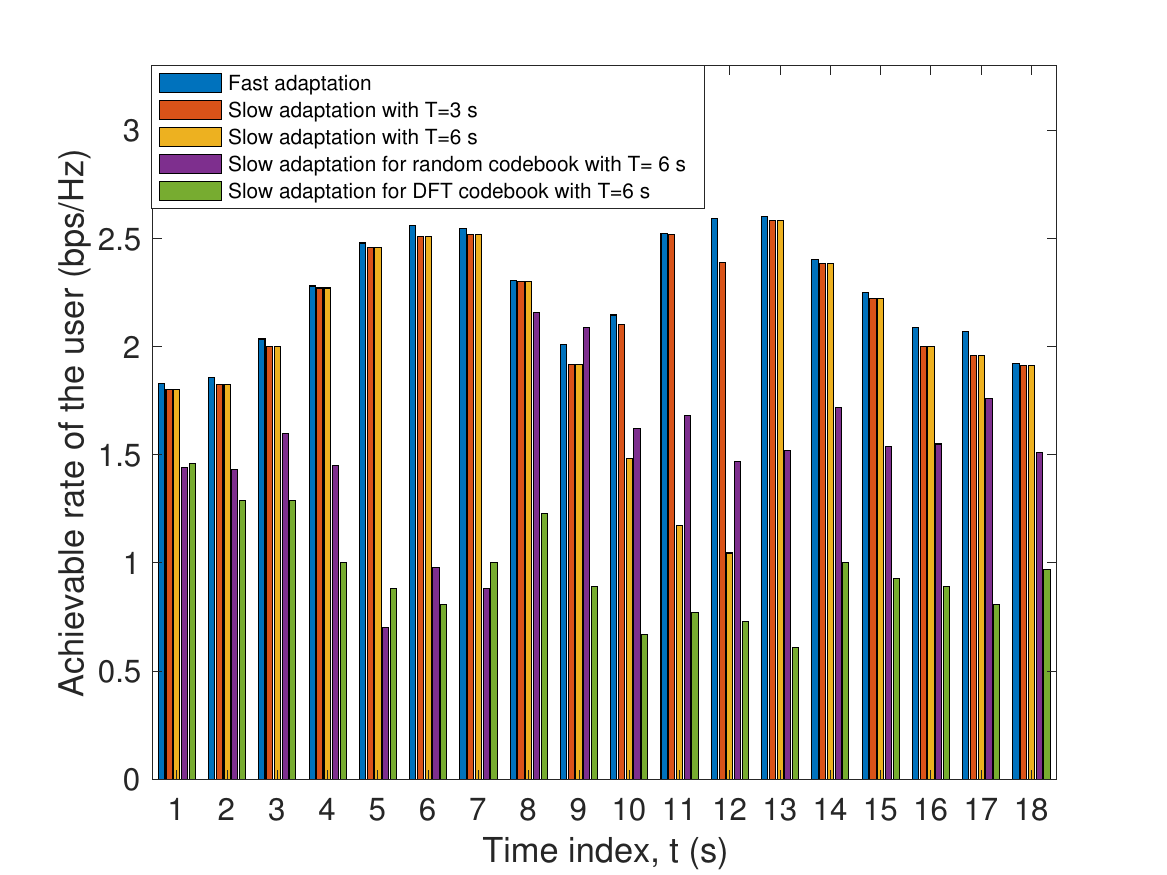}
		\end{minipage}
	}
    	\subfigure[$D=4$]{
    		\begin{minipage}[t]{0.45\textwidth}
   		 	\includegraphics[width=7cm]{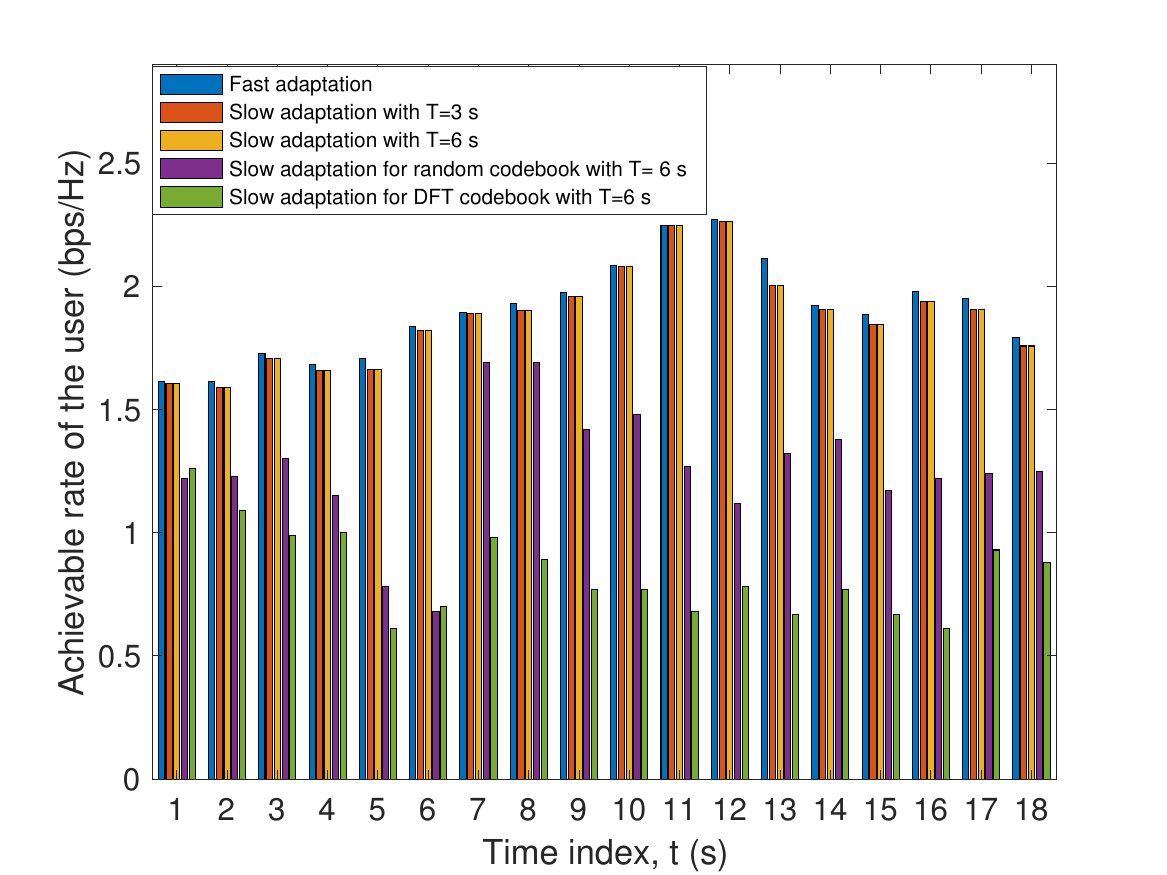}
    		\end{minipage}
    	}
	\caption{Performance comparison between fast adaptation and slow adaptation.}
	\label{fast_slow}
\end{figure}

\begin{figure}
	\centering
	\subfigure[$D=8$]{
		\begin{minipage}[t]{0.45\textwidth}
			\includegraphics[width=6.8cm]{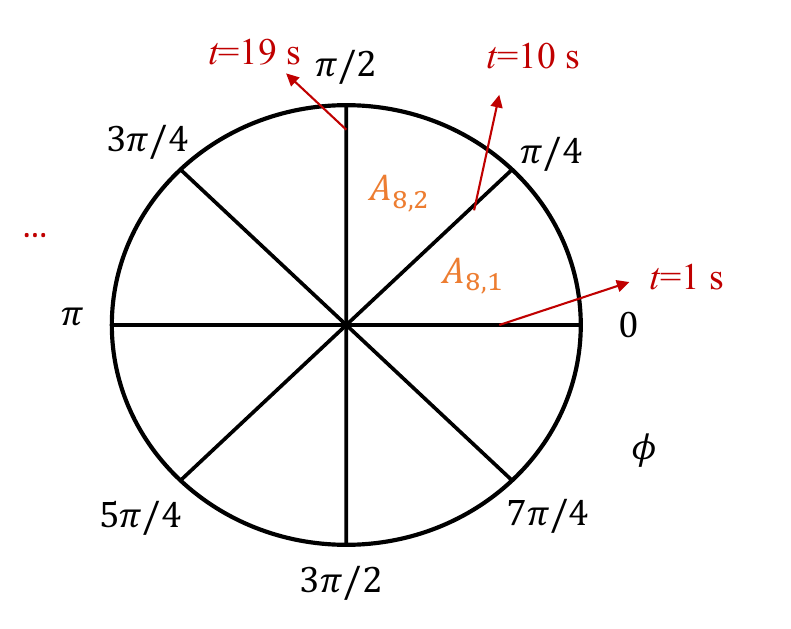}
		\end{minipage}
	}
    	\subfigure[$D=4$]{
    		\begin{minipage}[t]{0.45\textwidth}
   		 	\includegraphics[width=6.8cm]{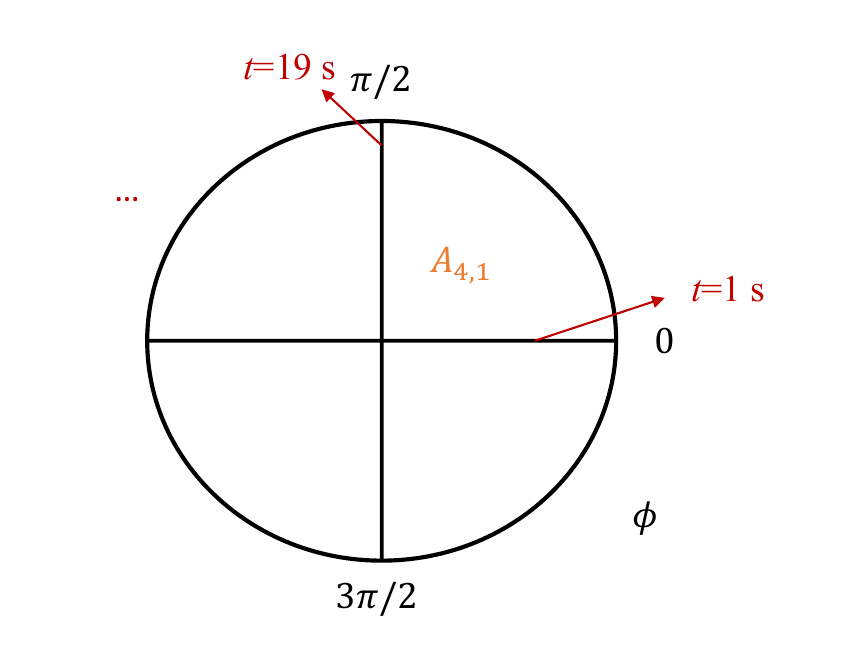}
    		\end{minipage}
    	}
	\caption{Illustration of user movement trajectory.}
	\label{user_movement}
\end{figure}

Fig. \ref{fast_slow} depicts the average achievable rate (over 400 fading blocks) at each time instant $t$ by respectively setting $D=8$ and $D=4$, where the Rician factor is set as $\kappa=10~\text{dB}$ and ``fast adaptation'' refers to updating the IRS reflection at each fading block, and Fig. \ref{user_movement} shows the corresponding user movement trajecotories. It is observed from Fig. \ref{fast_slow}(a) that in the case of $D=8$, the performance achieved by the slow adaptation with $T=3~\text{s}$ can approach that achieved by the fast adaptation at all time instants, while the performance achieved by the slow adaptation with $T=6~\text{s}$ becomes worse when $10~\text{s}\leq t\leq12~\text{s}$. This is because as shown in Fig. \ref{user_movement}(a), in the case of $D=8$, the user is in sector $\mathcal{A}_{8,1}$ when $1~\text{s}\leq t\leq 9~\text{s}$ and in sector $\mathcal{A}_{8,2}$ when $10~\text{s}\leq t\leq18~\text{s}$. Accordingly, for the slow adaptation with $T=3~\text{s}$, the IRS reflection pattern can always be updated to cover the sector where the user is located. In contrast, for the slow adaptation with $T=6~\text{s}$, the IRS reflection pattern is designed to cover sector $\mathcal{A}_{8,1}$ when $7~\text{s}\leq t\leq12~\text{s}$, while the user moves to sector $\mathcal {A}_{8,2}$ when $10~\text{s}\leq t\leq18~\text{s}$, which thus results in performance loss when $10~\text{s}\leq t\leq12~\text{s}$. Then, the IRS reflection pattern will be updated at $t=\text{13}~\text{s}$, which will cover $\mathcal{A}_{8,2}$ and re-boost the performance achieved by the slow adaptation with $T=6~\text{s}$ when $13~\text{s}\leq t\leq18~\text{s}$. The above results indicate that the time duration for fixing IRS reflection pattern, i.e., $T$, should be properly set based on the user's moving speed to achieve the best tradeoff between improving the user's achievable rate and reducing channel training overhead. 

Moreover, by comparing the performance achieved by the slow adaptation with $T=6~\text{s}$ in the case of $D=8$ (see Fig. \ref{fast_slow} (a)) and that in the case of $D=4$ (see Fig. \ref{fast_slow}(b)), it can be observed that the performance achieved by slow adaptation with $T=6~\text{s}$ is similar as that achieved by fast adaptation under all time instants in the latter case. Since in the latter case, when $1~\text{s}\leq t\leq18~\text{s}$, the user is always in sector $\mathcal{A}_{4,1}$ (see Fig. \ref{user_movement}(b)) and the IRS reflection always covers sector $\mathcal{A}_{4,1}$, which can ensure its average performance. This indicates that a smaller codebook size $D$ could help reduce the channel training overhead by setting a longer time duration for fixing the IRS reflection pattern, although it will compromise the user's achievable rate due to lower channel power gain of the designed codewords (see Fig. \ref{SIMO_D}).

In addition, it is observed from Fig. \ref{fast_slow} that the random codebook-based reflection design and DFT codebook-based reflection design cannot achieve effective slow adaptation as our proposed codebook-based reflection design, since the codewords in random codebook and DFT codebook cannot ensure the performance of all locations in each sector, which thus compromises the user's rate performance even though it is located in the same sector.

\subsection{Performance Evaluation for Multi-User Transmission}
\begin{figure}
\centering
\includegraphics[width=9cm]{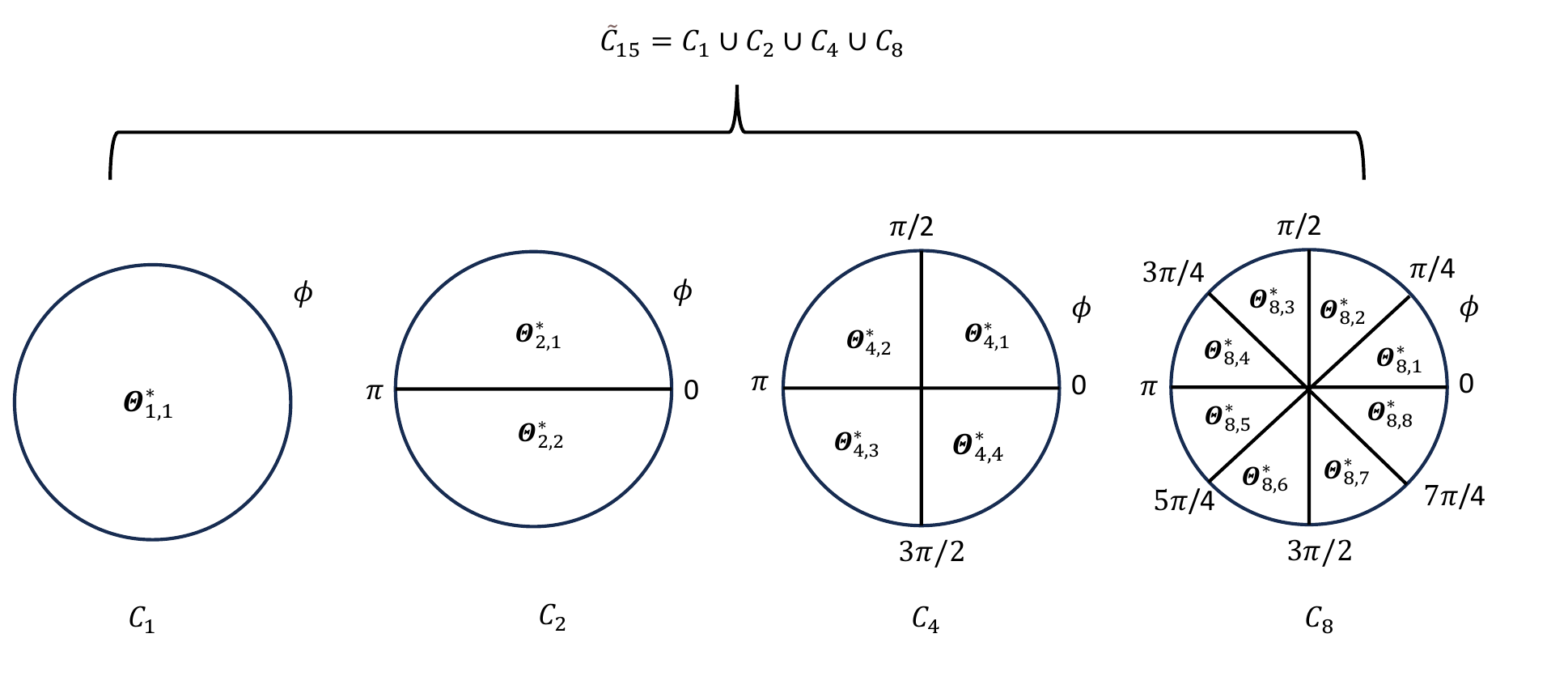}
\caption{Illustration of constructing $\tilde{\mathcal C}_{15}$.}\label{example_codebook}
\end{figure}

Finally, we discuss how to extend our proposed codebook-based IRS passive design for single-user transmission to multi-user transmission. Since in multi-user transmission, the users may be distributed at arbitrary locations in the coverage area $\mathcal A$, the IRS reflection should be designed to cover larger sectors including more users' locations in general, which thus leads to a small number of sectors for the selected codeword, $D$. As a result, how to select the value of $D$ for the codebook for multi-user transmission is a challenging problem in practice. To tackle this problem, we propose to construct the overall codebook for multi-user transmission as the union of all codebooks for single-user transmission with different values of $D$ (or numbers of sectors), where their sum is equal to the new codebook size for multi-user transmission, denoted as $X$. Let $\tilde{\mathcal{C}}_{X}$ denote the overall codebook for multi-user transmission with the size of $X$. For example, to construct the codebook $\tilde{\mathcal C}_{15}$, we first design the codebooks for single-user transmission with $D=1$, $D=2$, $D=4$, and $D=8$, respectively\footnote{Although we only consider the case with $D=1$, $D=2$, $D=4$, and $D=8$, to show the tradeoff between improving the performance and reducing channel training overhead, our proposed design can be extended to other arbitrary values of $D$ for multi-user transmission.}, and then construct $\tilde{\mathcal C}_{15}$ as the union of $\mathcal{C}_{1}$, $\mathcal {C}_{2}$, $\mathcal {C}_{4}$, and $\mathcal{C}_{8}$. Specifically, when $X=1+2+4+8=15$, we have $\tilde{\mathcal{C}}_{15}=\mathcal{C}_{1}\cup\mathcal{C}_{2}\cup\mathcal{C}_{4}\cup\mathcal{C}_{8}$. For ease of illustration, we show the construction of $\tilde{\mathcal C}_{15}$ in Fig. \ref{example_codebook}. The same procedure is applied for constructing the codebook $\tilde{\mathcal{C}}_{X}$ with other codebook sizes or values of $X$. 

To evaluate the performance of the above proposed codebook-based IRS reflection design for multi-user transmission, we set the number of users as $K=4$. The effective channel for each user $k$ is defined as $\mv h_{k}$, which can be obtained as (\ref{effective_channel_old}) and modeled as Rician fading (see Section \ref{simulation_results_section}-B). Similar as single-user transmission, the locations of the $K=4$ users are all randomly generated in the cell-edge of the coverage area $\mathcal{A}$ and the random-scattering environment is employed for generating NLoS channels. It is assumed that the IRS-AP adopts the minimum mean square error (MMSE) combining and successive interference cancellation (SIC) technique to decode the signals from different users, and thus the sum-rate of all users is given by $R_{sum}=\log_{2}\text{det}\Big(\mv I_{M}+\sum_{k=1}^{K}\frac{p_{k}}{\sigma^{2}}\mv h_{k}\mv h_{k}^{H}\Big)$ with $p_{k}=P/K$ denoting the transmit power of user $k$, which is used as the performance metric for simulation \cite{wireless_communication}.  

\begin{figure}
\centering
\includegraphics[width=7cm]{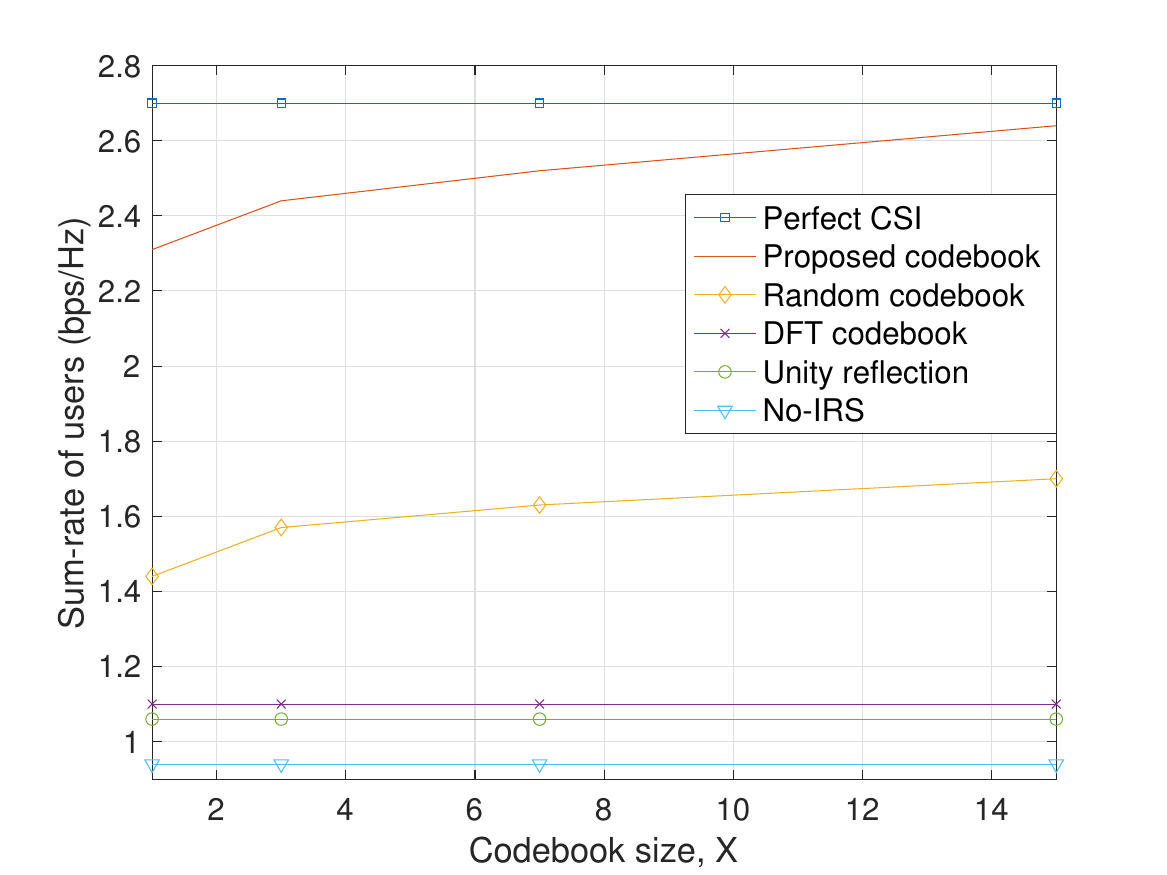}
\caption{Sum-rate of all users versus the codebook size $X$.}\label{rate_multi_user}
\end{figure}

Fig. \ref{rate_multi_user} shows the sum-rate of all users by all considered schemes versus the codebook size $X$, where we set $\kappa=10~\text{dB}$ and the codebook size for random codebook is also set as $X$. It is observed that our proposed codebook-based reflection design outperforms all benchmark schemes under all considered $X$ values in multi-user transmission. Specifically, by setting the codebook size as $X=15$, our proposed codebook-based reflection design is observed to achieve 55.29\%, 140.00\%, 149.06\%, and 180.85\% sum-rate improvements over random codebook, DFT codebook, unity reflection, and no-IRS schemes, respectively.

\begin{figure}
	\centering
	\subfigure[Single-user transmission]{
		\begin{minipage}[t]{0.45\textwidth}
			\includegraphics[width=7cm]{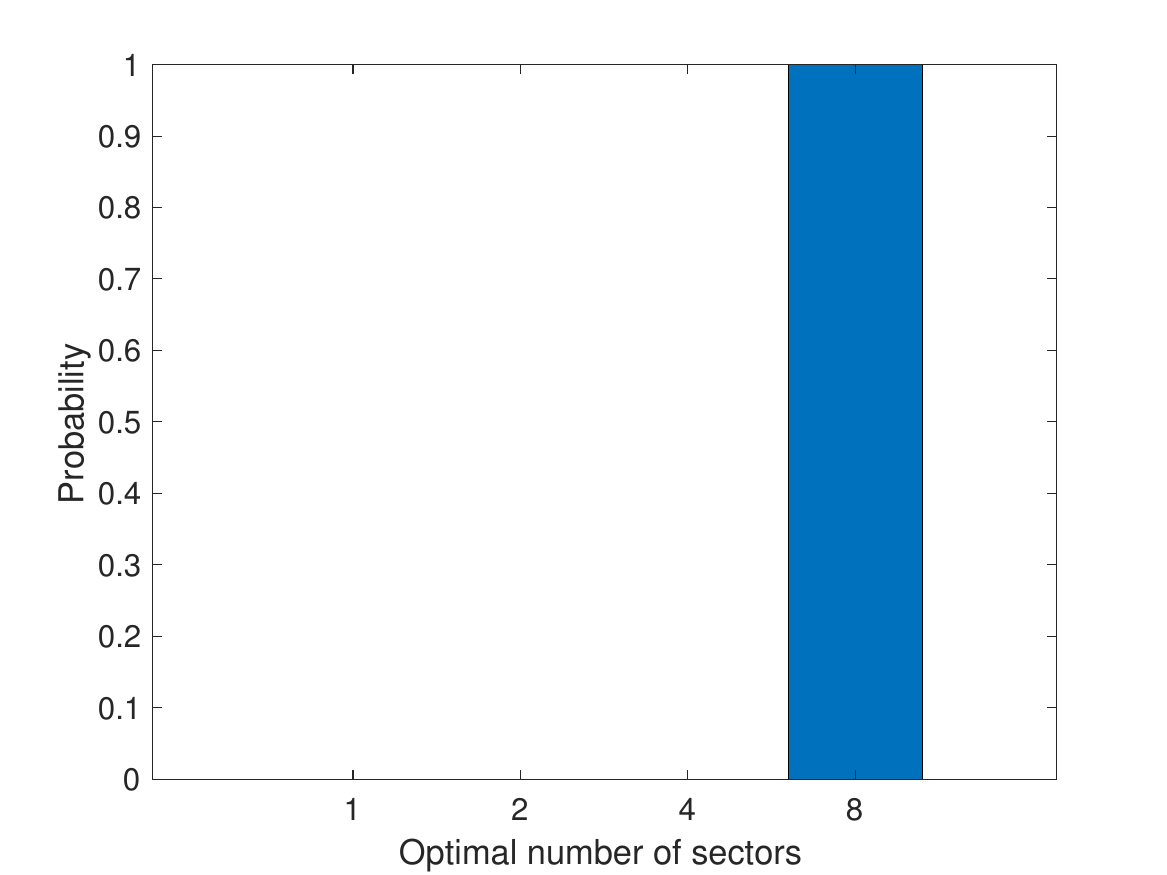}
		\end{minipage}
	}
    	\subfigure[Multi-user transmission]{
    		\begin{minipage}[t]{0.45\textwidth}
   		 	\includegraphics[width=7cm]{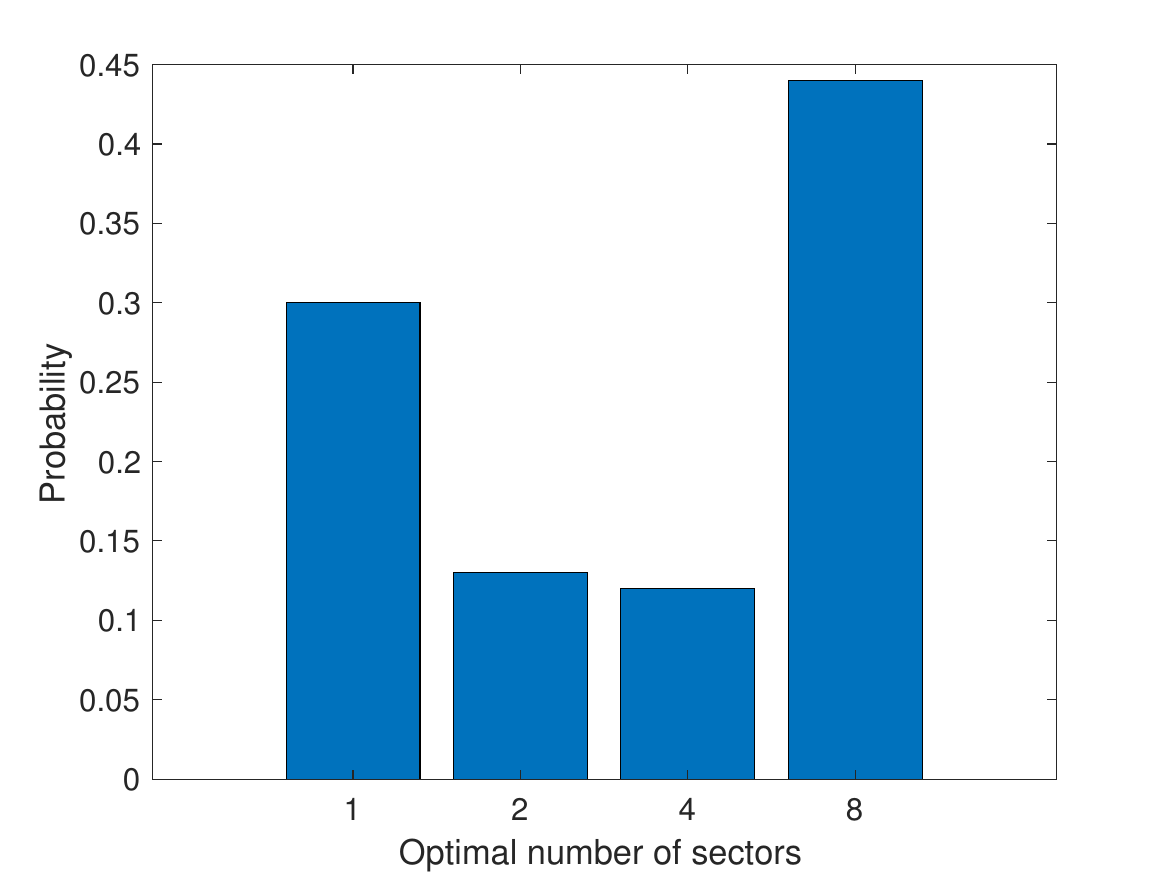}
    		\end{minipage}
    	}
	\caption{Probability of the optimal number of sectors for the selected codeword.}
	\label{probability}
\end{figure}

Moreover, Fig. \ref{probability} shows the probability of the optimal number of sectors for the selected codeword to achieve the best communication performance for single-user transmission (i.e., largest achievable rate of the user) and multi-user transmission (i.e., maximum sum-rate of the users), respectively, where we consider 1000 realizations of user locations and set $\kappa\rightarrow\infty$ for considering LoS environment. It is observed that the optimal number of sectors for single-user transmission is always fixed as its largest value (see Fig. \ref{probability}(a)), i.e., $8$, while that in multi-user transmission varies with different realizations of user locations (see Fig. \ref{probability}(b)). This is because in single-user transmission, the achievable rate of the user is only dependent on its effective channel power, which can be significantly enhanced by setting the number of sectors to the largest possible value to obtain the largest  power gain. In contrast, in multi-user transmission, the users are distributed at random locations in the coverage area $\mathcal{A}$, and the sum-rate of the users is affected by other factors, such as the interference among them, both of which render that the optimal number of sectors for the selected codeword varies with different user locations.

\section{Conclusions}
In this paper, we proposed a codebook-based IRS reflection design for the IRS-AP to enhance the average channel power of the whole coverage area. The codebook only contained a small number of codewords for achieving small codebook size to reduce the channel training overhead, which was designed offline based on the proposed sector division strategy. Then, each codeword for IRS reflection pattern was optimized to maximize the SMAECP in its served sector, which was efficiently obtained by adopting the AO and SDR methods. By applying the codewords in the designed codebook, the AP could select the optimal IRS reflection pattern with the best communication performance for data transmission. 

Numerical results demonstrated that our proposed codebook-based IRS reflection design can outperform other benchmark schemes in both single-user and multi-user transmissions. It was also shown that the proposed sector division strategy based on the azimuth angle enabled that the codebook-based IRS reflection can achieve slow adaptation to wireless channels, where the number of sectors and the time duration for updating IRS reflection should be properly selected to balance the tradeoff between improving user achievable rate and reducing channel training overhead. There are several promising directions worthy of further investigation for reflection codebook design in IRS-AP based communications, such as accounting for imperfections in IRS phase-shift model \cite{practical_ps} when designing the reflection codebook, designing the reflection codebook for frequency-selective wideband channels, designing the reflection codebook for MIMO systems, considering the reflection codebook design for NOMA systems, and so on. Moreover, more sophisticated/efficient algorithms for codeword design are worth further studying. Furthermore, the IRS reflection codebook for active IRS \cite{active_irs1,active_irs2} is also an interesting direction to pursue. Last but not least,  when and how to take more than two-hop reflections into consideration \cite{multi_hop} for codebook design also needs to be further investigated.

\end{document}